\documentclass[%
 reprint,
 amsmath,amssymb,
 aps, pre,
]{revtex4-2}

\usepackage{graphicx}
\usepackage{dcolumn}
\usepackage{bm}

\usepackage{enumerate}

\begin{document}


\title{Ionization/dissociation-driven first order phase transitions: \\
on a new class of first order phase transitions}

\author{Genri Norman$^{1,2}$}
\author{Ilnur Saitov$^{3,1}$}%
 \email{ilnur.saitov@univaq.it}

\affiliation{$^1$Joint Institute for High Temperatures of RAS, Izhorskaya st. 13, Bld.2, Moscow,125412, Russia}
\affiliation{$^2$Moscow Institute of Physics and Technology (National Research University), Institutskiy per. 9, Dolgoprudny, Moscow Region, 141700, Russia}
\affiliation{$^3$Department of Physical and Chemical Sciences, University of L’Aquila, L'Aquila, 67100, Italy}

\date{\today}

\begin{abstract}
In this work, we compare various models for describing the phase transition of the fluid hydrogen into a conducting state, including both chemical models of plasma and first-principle simulations within the framework of the density functional theory (DFT). The comparison of the results indicates the plasma nature of the phase transition in warm dense hydrogen. We propose a concept of a new class of first-order phase transitions: ionization or dissociation-driven phase transitions, with which the plasma phase transition in fluid hydrogen can be associated.
\end{abstract}

\maketitle
The plasma phase transition (PPT) was suggested in 1968  \cite{Norman1968}. First, the research was limited to theoretical works \cite{Biberman1969,Norman1970,Norman1970a,Ebeling1969,Ebeling1971,Insepov1972,Filinov1975,Ebeling1985,Ebeling1985a,Kraeft1986,Saumon1989,Saumon1991,Saumon1992,Saumon1995,Reinholz1995,Norman2000,Norman2001,Ebeling2003,Scandolo2003,Bonev2004,Bonev2004a,Gryaznov2004}. In 2007, the first experimental indication of a phase transition in warm dense hydrogen at megabar pressures appeared \cite{Fortov2007}. Since then, a significant number of studies have been published on this topic, both theoretical \cite{Vorberger2007,Gryaznov2009,Tamblyn2010,Tamblyn2010a,Lorenzen2010,Gryaznov2010,Morales2010,Gryaznov2013,Khomkin2013,Morales2013,McMinis2015,Gryaznov2015,Starostin2016,PierleoniMoralesRilloEtAl2016,Pierleoni2017,Rillo2018,Mazzola2018,Geng2019,Cheng2020,Hinz2020,Gorelov2020,Fedorov2020,Karasiev2021,Fedorov2021,bonitz2024toward,tenti2024principal,tenti2025hydrogen,tirelli2022high,fried2022thermodynamic,longo2023interfacial,buldyrev2024monte,Buldyrev2024,bergermann2024nonmetal,istas2025liquid,lue2024re,dharma2025ionic} and experimental \cite{Loubeyre2012,Dzyabura2013,Ohta2015,Knudson2015,McWilliams2016,Zaghoo2016,Zaghoo2017,Zaghoo2018,Celliers2018,Jiang2020}. A large number of studies focus on the phase diagram of dense hydrogen. However, the main attention is paid to the phase diagram in pressure ($P$) – temperature ($T$) coordinates. 

In the given paper, we consider the phase diagram for the same phase transition in pressure–specific volume ($P$–$V$) coordinates. Within both chemical models and the \textit{ab initio} quantum approach, three features of the $P(V)$ isotherms are identified, distinguishing these isotherms from van der Waals–type $P(V)$ isotherms.

The first section explains the essence of the differences from van der Waals–type $P(V)$ isotherms discussed in the paper. The second, third, and fourth sections present examples from several chemical models. The fourth section focuses on the phase diagram obtained within the ab initio quantum molecular dynamics approach \footnote{This refers to molecular dynamics simulations using density functional theory to compute interparticle forces.}. The sixth section provides arguments in support of the plasma nature of the fluid–fluid phase transition in warm dense hydrogen. The seventh section summarizes the conclusions.

\section{The essence of the issue}
\begin{figure}
    \centering
    \includegraphics[width=1\linewidth]{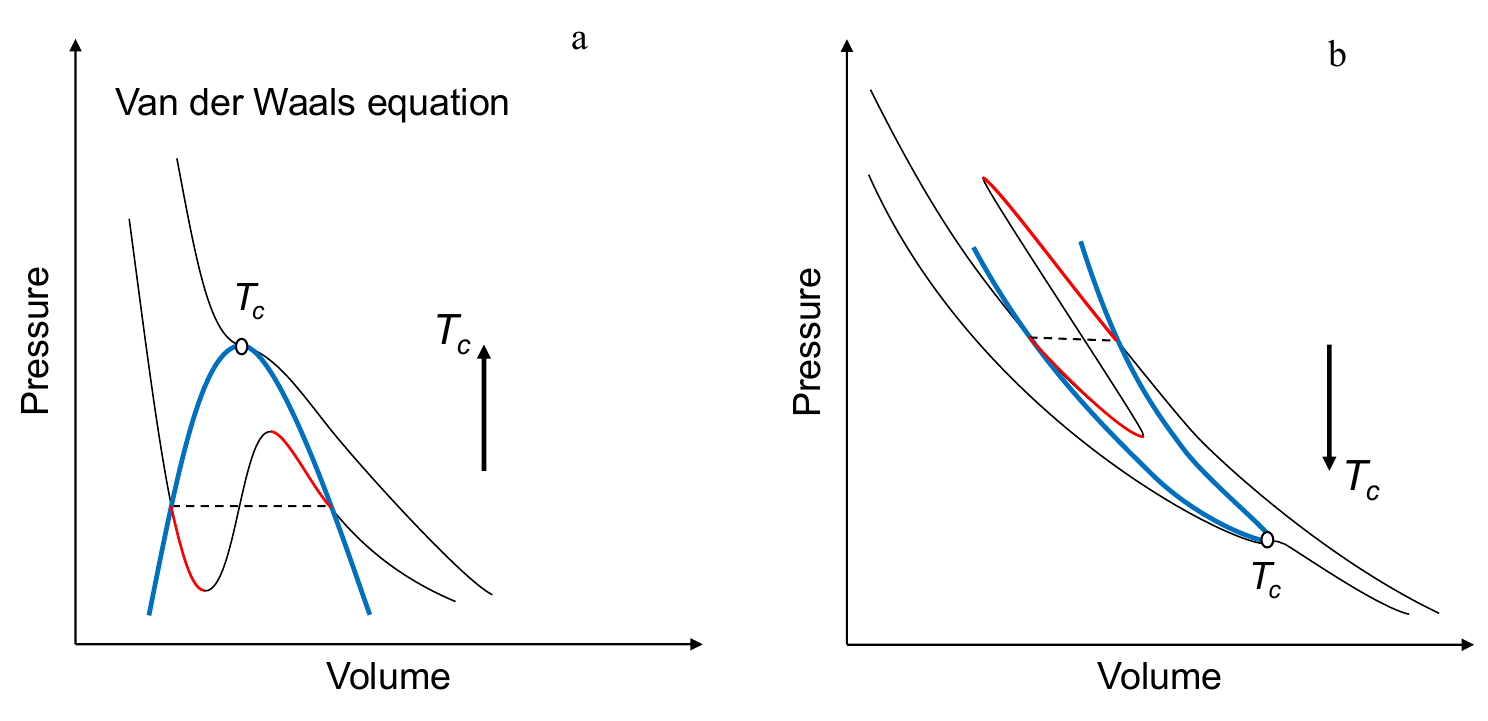}
    \caption{Schematic representation of the $P(V)$ isotherm (black lines) and the binodal (blue line) for the van der Waals gas (a) and the considered phase transition in hydrogen (b). The red segments represent metastable states. The dashed line indicates the phase equilibrium pressure. The critical point $T_c$ is shown.}
    \label{fig01}
\end{figure}

We illustrate the essence of the issue in Fig.~\ref{fig01} \cite{Norman2021}, which presents two types of $P(V)$ isotherms and schematically shows three features of the $P(V)$ isotherms that distinguish the considered phase transitions from the usual picture given by the van der Waals equation.

1. On the left, the function $P(V)$ is single-valued; therefore, the metastable and equilibrium branches of different phases are separated across distinct ranges of specific volume. On the right, an overlap is observed in a certain range of specific volume between the metastable branch of one phase and the metastable and equilibrium branches of the other phase.

2. The critical temperature corresponds to the highest pressure among all isotherms on the left, and to the lowest pressure on the right
\footnote{The first indication of this property of the plasma phase transition was given in Ref.~\cite{Ebeling1985a}, where the $P$–$T$ coordinates were considered.}.

3.	Due to the three-valued nature of $P(V)$, there is an isolated region of metastable states on the right.

\begin{figure}
    \centering
    \includegraphics[width=0.8\linewidth]{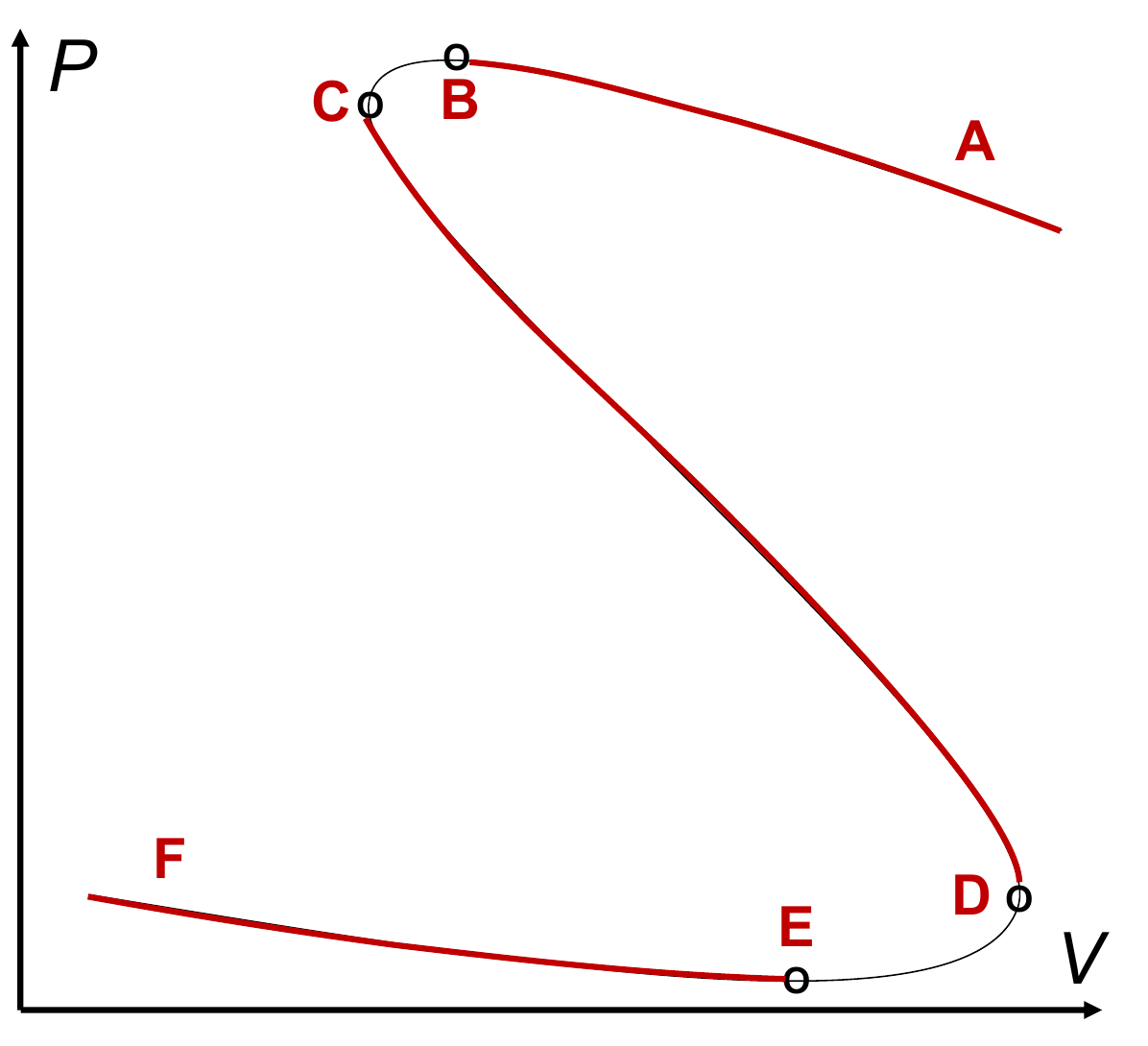}
    \caption{Schematic representation of the S-shaped, three-valued segment of the $P(V)$ isotherm.
Points B and E are the spinodal points; segments AB and EF are metastable regions adjacent to the equilibrium segments (not shown in the figure) on the left and right, respectively; BC and DE are unstable (labile) regions; CD is an isolated segment of metastable states.}
    \label{fig02}
\end{figure}

The last feature is not entirely obvious from Fig.~\ref{fig01} due to the small scale, so we isolate the S-shaped segment of the $P(V)$ isotherm between the two metastable branches in a separate Fig.~\ref{fig02}. From Fig.~\ref{fig02}, one can see that a new, isolated region of metastable states has appeared between the two unstable segments BC and DE.

In Figs.~\ref{fig01} and \ref{fig02}, we have schematically summarized and previewed the results for several specific models in advance. We provide this preview due to the unusual nature of these results. Now we proceed to a systematic presentation of the main results for each of the models.

\section{Plasma Phase Transition in the Simplest Chemical Model of a Three-Component Plasma}
The chemical model is an approximation in which the system is initially divided into atoms, molecules, atomic and molecular ions, and electrons. The concentrations of each species are determined by solving the set of ionization and chemical equilibrium equations, taking into account interactions between particles. In this section, as well as in the Sections III and IV, we briefly describe known examples of such calculations for phase transitions of the type we are interested in.

\subsection{Historical background}
The first example goes back to the 1968 work \cite{Norman1968}. There, the simplest chemical model was considered: (a) atoms were treated as an ideal gas, (b) long-range effective Coulomb attraction between charges was taken into account in the Debye–Hückel approximation, (c) short-range quantum repulsion of electrons and ions was included via the second virial coefficient in the Vedenov–Larkin approximation. The corresponding expression for the (b) and (c) contributions to the free energy is
\begin{equation}\label{01}
	\frac{{\Delta F}}{{nk_{B}T}} = - \frac{2}{3}{\pi^{1/2}}{\gamma^{3/2}}\left( {1 - 0.075\lambda \kappa } \right),
\end{equation}
where $\gamma = {e}^{2}{{n}{e}}^{1/3}/k_{B}T$ is the nonideality parameter, $n = n_e + n_i = 2n_i$ is the total concentration of electrons and singly charged ions, $\lambda$ is the electron de Broglie wavelength, and $\kappa^{-1}$ is the Debye radius.

The phase transition and critical point are the result of the competition between the long-range effective Coulomb attraction, the short-range effective quantum repulsion of electrons from ions, and temperature. Conceptually, it is analogous to the classical picture of the van der Waals equation, where the phase transition and critical point arise from the competition between a weak long-range attraction, a strong short-range repulsion, and temperature. At the same time, we have to impose stability criteria.

\subsection{Thermodynamic stability of plasma}
The criterion of thermodynamic stability, $(dP/dV)_T < 0$, is reduced in Ref.~\cite{Norman1968} for the considered chemical model to the inequality
\begin{equation}\label{02}
	\left( \partial n_a/\partial n_i\right)\geq 0.
\end{equation}
Therefore, thermodynamic stability can be analyzed using the ionization equilibrium equation, which for the considered chemical model can be expressed as follows
\begin{equation}\label{03}
\begin{gathered}
	{n_a} = n_i^2\frac{{{\Sigma _a}}}{{2{\Sigma _i}}}{\left( {\frac{{2\pi {\hbar ^2}}}{{{m_e}{k_B}T}}} \right)^{3/2}}\times \\ 
  \times \exp \left[ {\frac{I}{{{k_B}T}} - 2{\pi ^{1/2}}{\gamma ^{3/2}}\left( {1 - 0.1\lambda \kappa } \right)} \right],
\end{gathered}
\end{equation}  
where $\Sigma_a$ and $\Sigma_i$ are partition functions of atoms and ions respectively, $I$ is the isolated atom ionization energy. 

\begin{figure}[]
    \centering
    \includegraphics[width=0.8\linewidth]{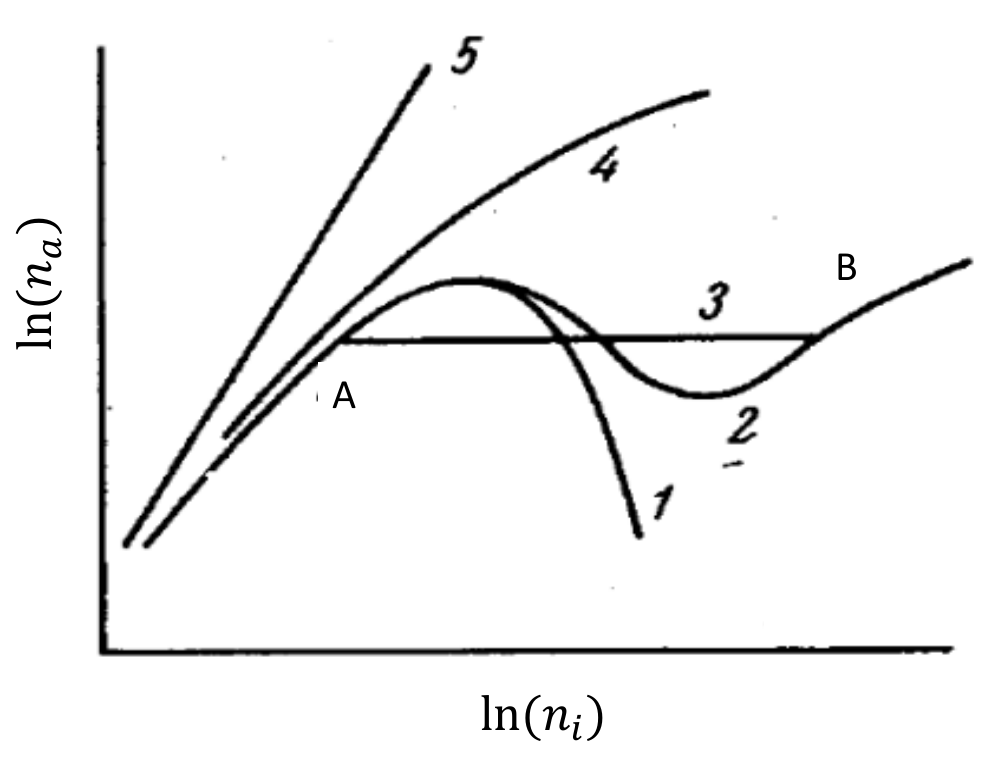}
    \caption{Dependencies of atomic concentration on concentration of ions at constant temperature, obtained under various approximations: 1 is the case of a Debye plasma; 2–4 are possible shapes of the curve taking into account the quantum nature of the system (in the region near point A and to the left, curves 1, 2, and 3 are assumed to coincide for illustrative purposes; to the right of point B, curves 2 and 3 are also assumed to coincide); 5 is an ideal gas of electrons and ions (this dependence comes from the formula (\ref{03}) without ionization potential lowering).}
    \label{fig03}
\end{figure}

The various dependencies $n_a(n_i)$ along the isotherms considered in \cite{Norman1968} are shown in Fig.~\ref{fig03}.

The variant 1, which provides the loss of stability, in particular prompted the authors of Ref.~\cite{Norman1968} to search for a factor that would ensure the stability. This factor turned out to be the short-range effective quantum repulsion of electrons from ions.

\begin{figure}[]
    \centering
    \includegraphics[width=0.8\linewidth]{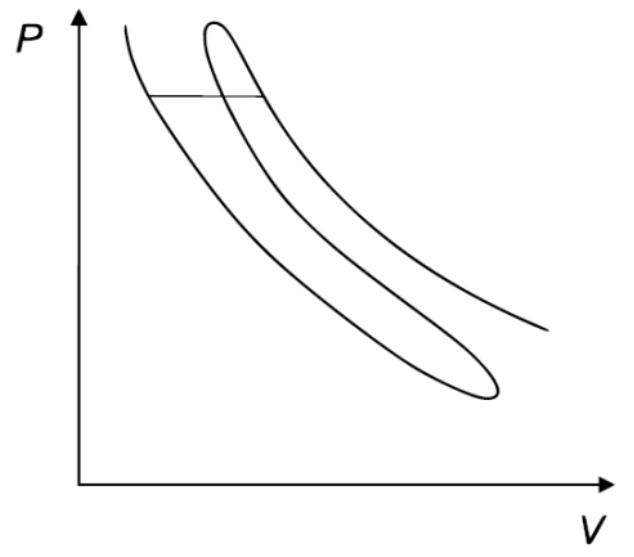}
    \caption{The $P(V)$ isotherm for the case where the inequality (\ref{05}) is satisfied \cite{Biberman1969}.}
    \label{fig04}
\end{figure}

The following relation between $(\partial P/\partial V)_{T}$ and $(\partial n_a /\partial n_i)_{T}$ has been obtained in Ref.~\cite{Norman1968}
\begin{equation}\label{04}
\begin{gathered}
    \left(\frac{\partial P}{\partial V}\right)_{T,{N_i},{N_a}} = -\frac{{n_a} kT}{V^2}\left({\frac{{\partial {n_a}}}{{\partial {n_i}}}}\right)_T \times \\ \times\left(1 +  \frac{n_i}{n_a}\right)^{2}\left[ 1 + \left(\frac{\partial {n_a}}{\partial {n_i}} \right)_T \right]^{-1},
\end{gathered}
\end{equation}
where $N_{a}$ and $N_{i}$ are the numbers of atoms and ions, respectively. The authors of article \cite{Norman1968} paid attention to the multiplier $(\partial n_{a} /\partial n_{i})_{T}$ on the right-hand side and derived the stability condition (\ref{02}). However, they did not notice that another stability condition follows from the last multiplier in (\ref{04}).
\begin{equation}\label{05}
	\left(\partial n_{a}/\partial n_{i}\right)_{T} < -1.
\end{equation}
The condition (\ref{05}) was noted in a subsequent work \cite{Biberman1969}, where, for the first time, the existence of an isolated region of metastable states was also pointed out. Until 1969, both the van der Waals loop and all other $P(V)$ loops presented in the literature were single-valued functions. The equation of state $P(V)$ shown in Fig.~\ref{fig04} was predicted in Ref.~\cite{Biberman1969} without performing any calculations, based solely on formula (\ref{05}). As far as we know, this is the first indication of the possibility that, within a certain range of specific volume, the metastable branch of $P(V)$ for one phase can overlap with the metastable and equilibrium branch of $P(V)$ for another phase.

Going back to Fig.~\ref{fig02}, we can conclude that the stability along segments BA and FE is provided by the fulfillment of criterion (\ref{02}), while for the segment CD it is ensured by the condition (\ref{05}). Thus, the physical meaning of the existence of the isolated region of metastable states is associated with a very steep decrease in $n_a$ as $n_i$ increases in this region.

Various aspects of the isolated region of metastable states were later examined in Refs.~\cite{Norman2000,Norman2001,Norman2001rydberg,Norman2006}. Attempts to relate this to the phenomenon of ball lightning were made in Ref.~\cite{Biberman1969,Norman2000}.

\subsection{Padé approximation model}
Returning to the chemical model \cite{Norman1968}, we rewrite the equation (\ref{01}) in a shorter form
\begin{equation}\label{06}
	\frac{{\Delta F}}{{nk_{B}T}} = - \frac{2}{3}{\pi^{1/2}}{\gamma^{3/2}}\left( {1 - C\lambda \kappa } \right).
\end{equation}
As it is noted in Ref.~\cite{Norman1970a}, the following expression is valid as well within the first-order accuracy
\begin{equation}\label{07}
  \frac{{\Delta F}}{{nk_{B}T}} =  - \frac{2}{3}{\pi ^{1/2}}{\gamma ^{3/2}}\frac{1}{{1 + C\lambda \kappa }}.
\end{equation}
The expression (\ref{07}) is applied in the theory of electrolytes. However, the formulas (\ref{06}) and (\ref{07}) provide quite different results for plasma. Therefore, for methodological purposes, we use the Padé approximation model \cite{Ebeling1976theory,Kraeft1986} in its simplest form
\begin{equation}\label{08}
  \frac{{\Delta F}}{{nkT}} =  - \frac{2}{3}{\pi ^{1/2}}{\gamma ^{3/2}}\frac{{1 - \alpha C\lambda \kappa }}{{1 + \beta C\lambda \kappa }},
\end{equation}
where $\alpha + \beta = 1$ and $C = 0.075$. From the expression (\ref{08}), the ionization equilibrium equation follows
\begin{equation}\label{09}
  n_a = \frac{1}{4\sqrt \pi}{\left( {\frac{{k_BT}}{{{e^2}}}} \right)^3}{x^6}{\left( {\frac{{k_BT}}{{Ry}}} \right)^{3/2}}\exp \left[ {\frac{I}{{k_BT}} - \varphi } \right],
\end{equation}
where $ x = {\pi^{1/3}}\gamma $, $Ry = 13.6$~eV is the Rydberg constant. For methodological calculations, we chose the value $\alpha = 0.4$ (the effect of interest occurs for $\alpha < 0.5$) and $I=3.89$\,eV. The value $\varphi$ is the lowering of the ionization energy, which is determined as the derivative $\varphi = -2(V/k_BT)(\partial\Delta F/\partial N_i)$. 

\begin{figure}[]
  \centering
  \includegraphics[width=1.0\linewidth]{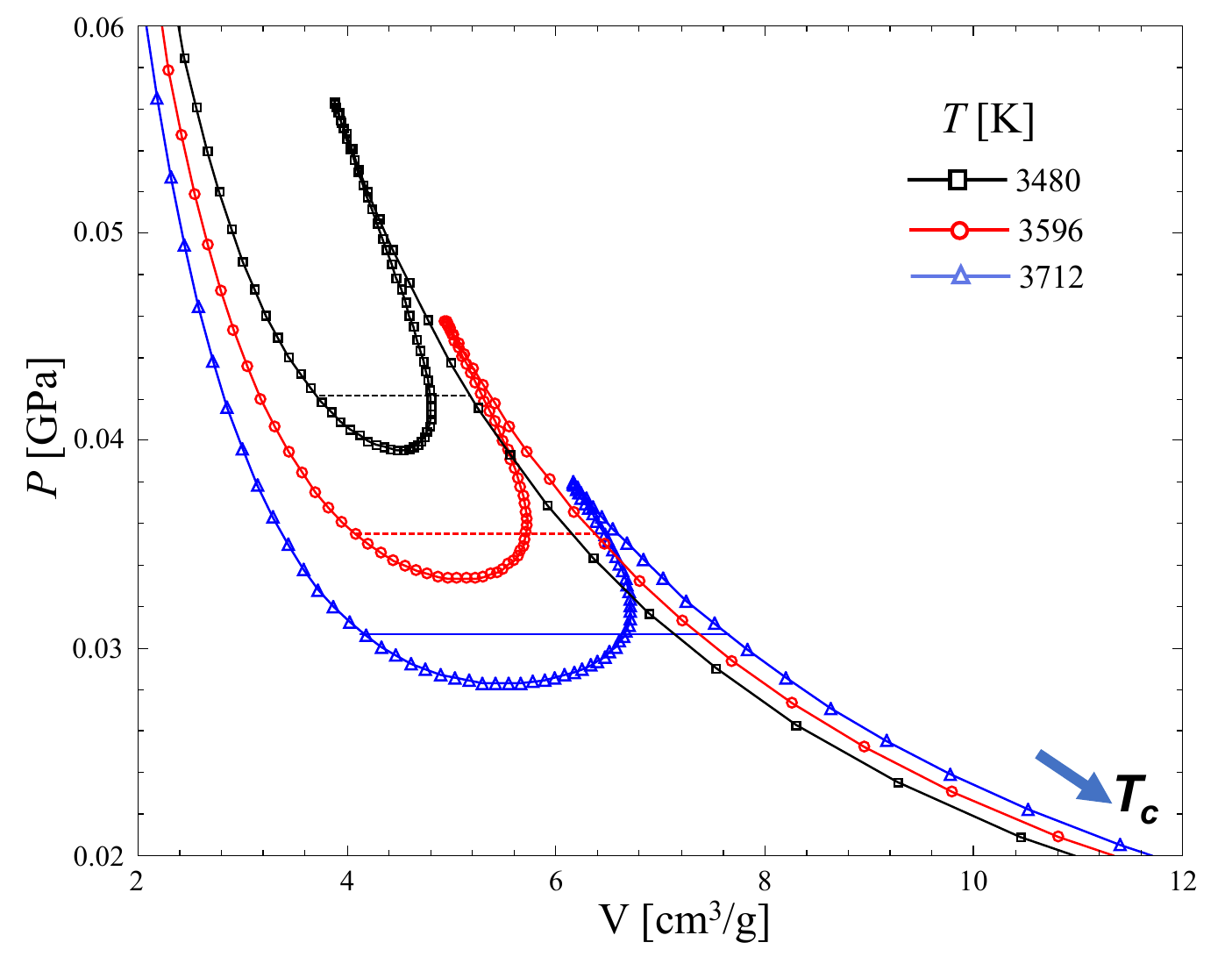}
  \caption{The equation of state of the Padé approximation model (\ref{10}) at three values of temperature. The arrow indicates the direction of increasing temperature as the system approaches the critical point}
  \label{fig05}	
\end{figure}

From the expressions (\ref{08}) and (\ref{09}) the following equation of state is derived 
\begin{widetext}
\begin{equation}
P = {n_a}kT + nkT\left[ {1 - \frac{1}{3}{x^{3/2}}\left( {1 - \frac{{1 + {C_1}{{\left( {kT/Ry} \right)}^{1/2}}{x^{3/2}}\left( {\beta  - \alpha  - 1} \right) + 5\alpha \beta C_1^2\left( {kT/Ry} \right){x^3}}}{{{{\left( {1 + \beta {C_1}{{\left( {kT/Ry} \right)}^{1/2}}{x^{3/2}}} \right)}^2}}}} \right)} \right],
 \label{10}
\end{equation}
\end{widetext}
where $C_1 = 0.385$. We show the isotherms $P(V)$ in Fig.~\ref{fig05} computed with the expressions (\ref{09}) and (\ref{10}) at the three values of temperature. As one can see, the results obtained confirm all three features of the $P(V)$ isotherms discussed above.

\begin{figure}[]
  \centering
  \includegraphics[width=1.0\linewidth]{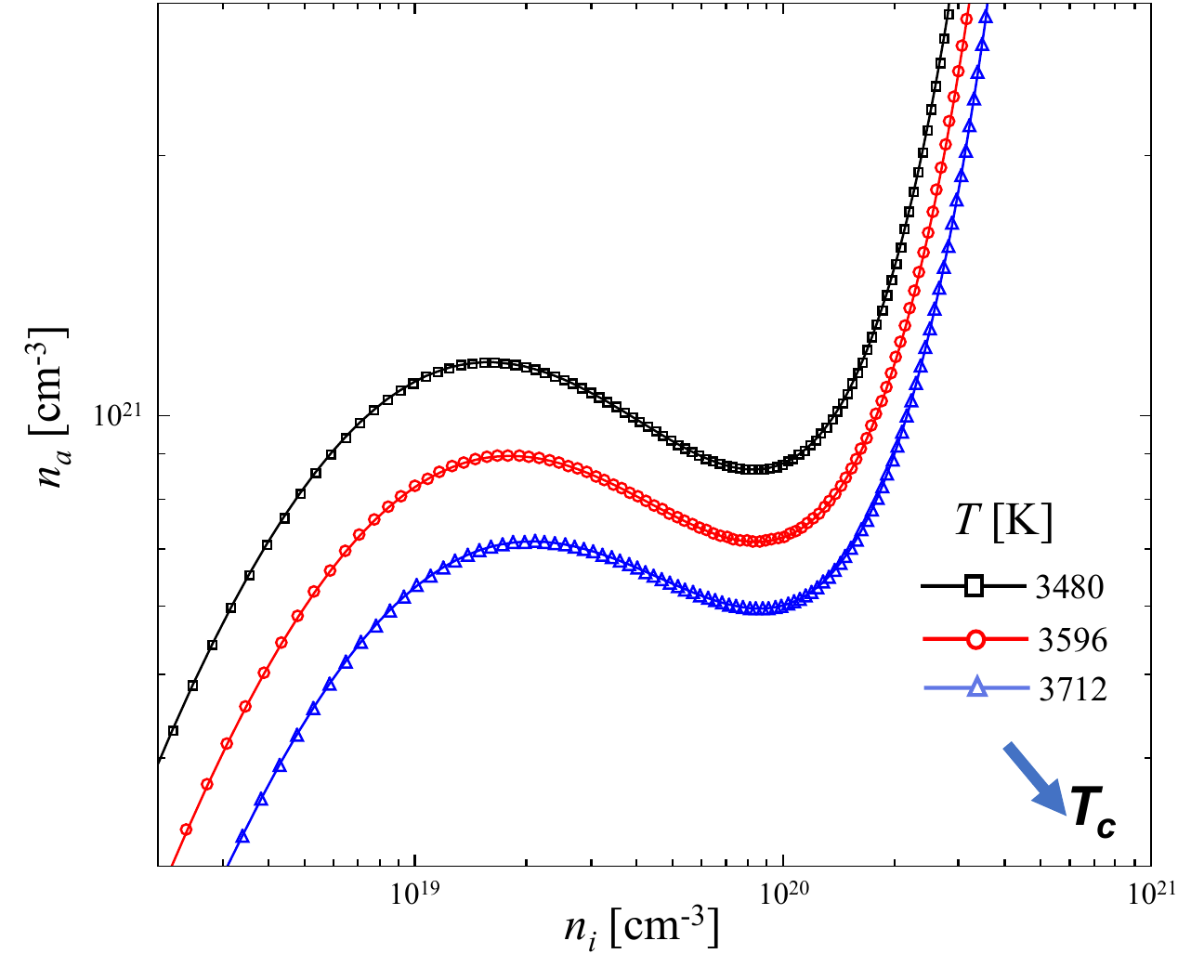}
  \caption{Dependence of atomic concentration on ion concentration at three different temperatures}
  \label{fig06}	
\end{figure}

For the same temperatures as in Fig.~\ref{fig05}, we have calculated the dependencies of atomic concentration on ion concentration using the formula (\ref{09}). The results are shown in Fig.~\ref{fig06}.

\section{Plasma phase transition in the chemical model of multicomponent plasma}
Chemical models of multicomponent plasma have been considered in a number of works \cite{Ebeling1969,Ebeling1971,Ebeling1985,Ebeling1985a,Saumon1989,Saumon1991,Saumon1992,Saumon1995,Reinholz1995,Fortov2003,Gryaznov2004,Gryaznov2009,Gryaznov2010,Gryaznov2013,Gryaznov2015,Starostin2016}, in which, on the one hand, new components beyond electrons and ions have been included, and interactions between all components have been taken into account. On the other hand, the treatment of these interactions has been progressively refined.

Let us focus on a more recent and the most advanced high-level chemical model: SAHA-D \cite{Gryaznov2009}. In this model, the hydrogen plasma is described as an equilibrium mixture of hydrogen atoms, molecules, atomic and molecular ions H, H$_2$, H$^+$, H$_2^+$, and electrons, interacting with each other, with electrons possibly being partially degenerate. In the case of deuterium, the components are D atoms, D$_2$ molecules, D$^+$ ions, D$_2^+$ molecular ions, and electrons. The model is five-component.

\begin{figure}[]
  \centering
  \includegraphics[width=1.0\linewidth]{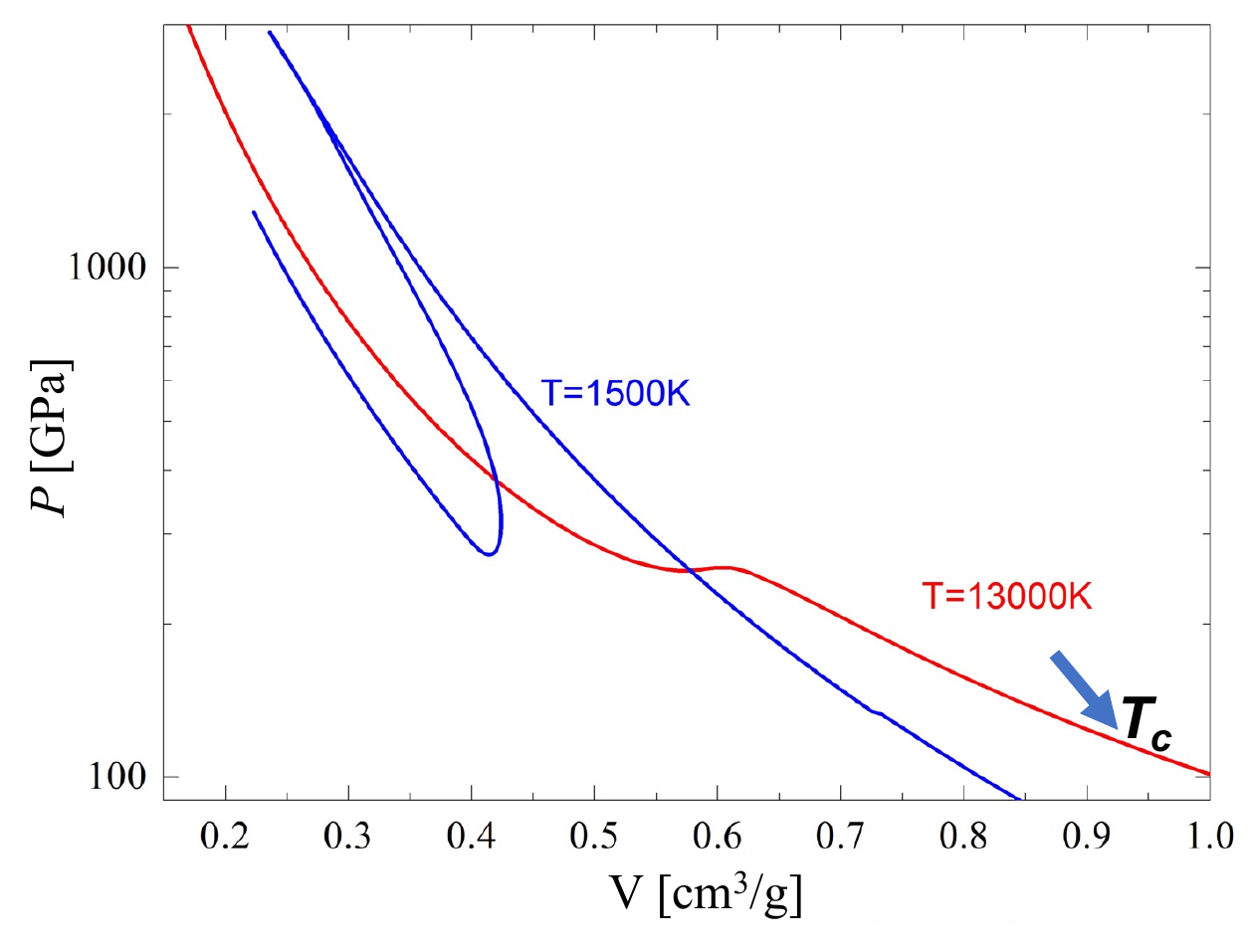}
  \caption{Equation of state $P(V)$ of the warm dense deuterium at two temperatures within the SAHA-D model \footnote{We thank V.K. Gryaznov for kindly providing us from his personal archive with this figure, which has not been previously published.}}
  \label{fig07}	
\end{figure}

The results of calculations of the equation of state $P(V)$ for the warm dense deuterium in the SAHA-D model for a subcritical temperature of 1500~K and a near-critical temperature of 13000~K are shown in Fig.~\ref{fig07}. The five-component model \cite{Gryaznov2009} differs significantly from the three-component model considered in the previous section due to the inclusion of the molecules and the molecular ions. It leads to the origin of new types of dissociation equilibria, which are absent in the three-component model.

However, the dependencies $P(V)$ in Fig.~\ref{fig05} and in Fig.~\ref{fig07} are quite similar to each other. The same sharp, elongated ''beak'' for the upper half-wave and a smooth, broad lower half-wave are observed. The phase equilibrium pressure is not marked in Fig.~\ref{fig07}, but it is clear that it decreases with increasing temperature as the system approaches the critical point. In a certain range of specific volume, the metastable branch of one phase overlaps with the metastable and equilibrium branches of another phase. Due to the three-valued nature of $P(V)$, an intermediate isolated region of metastable states exists. So, we can conclude that Coulomb interaction, which is common to both models, is the dominant factor.

Seven years after the publication of the article \cite{Gryaznov2009}, the current version of the SAHA-D chemical model was applied to hydrogen plasma in Ref.~\cite{Starostin2016}. The nature of two already observed phase transitions was analyzed.

\begin{figure}[]
  \centering
  \includegraphics[width=1.0\linewidth]{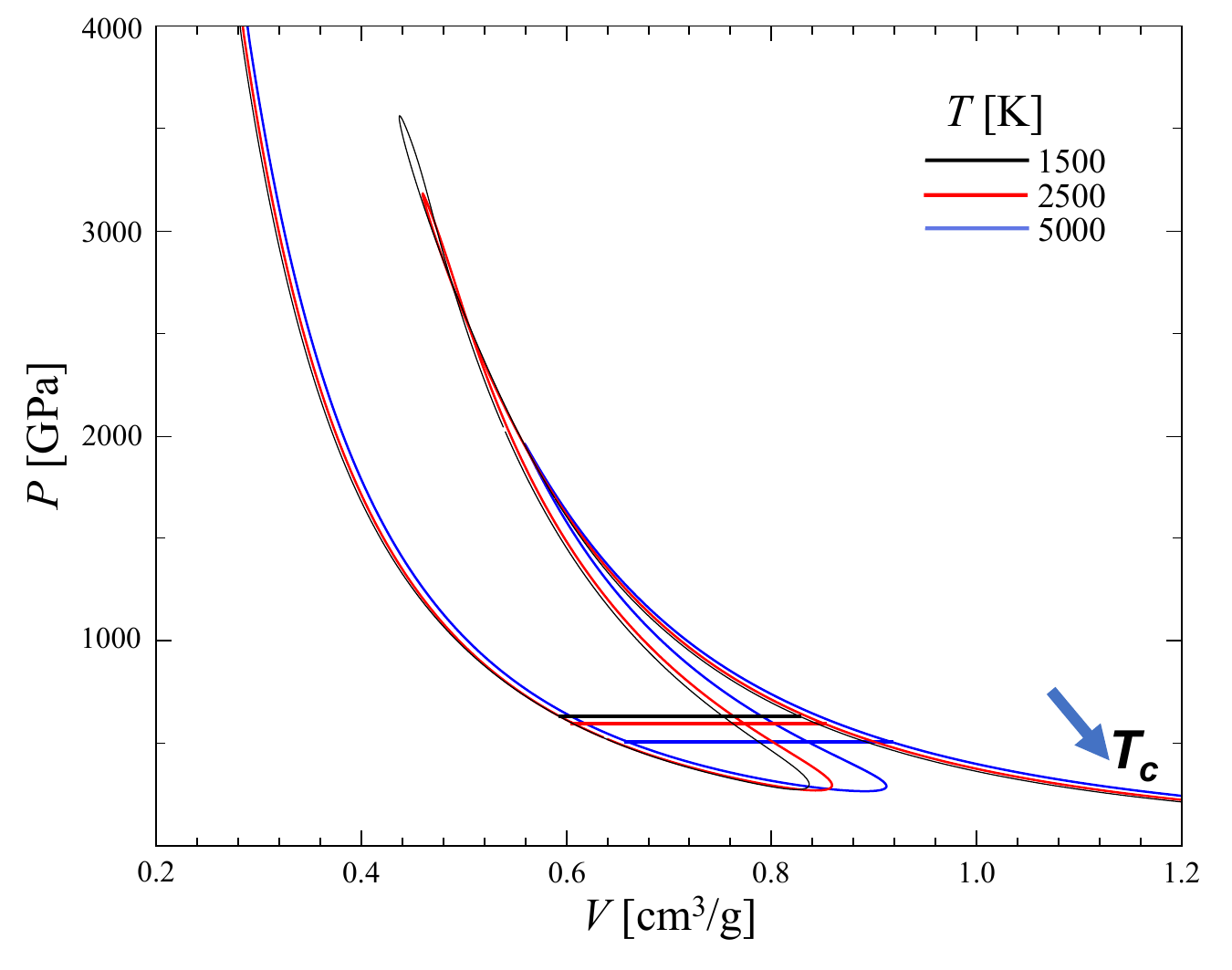}
  \caption{The equation of state $P(V)$ and the phase transition associated with the change of the major neutral plasma component from H$_2$ to H for three temperatures in the current version of the SAHA-D model \cite{Starostin2016}. \footnote{We thank A.V. Filippov for kindly providing us from his personal archive with this figure, which has not been previously published}}
  \label{fig08}	
\end{figure}

The results for relatively small specific volumes at $T=$ 1500, 2500, and 5000~K are shown in Fig.~\ref{fig08}. The model in Ref.~\cite{Starostin2016} differs significantly from the model described by the expression (\ref{08}). The physical nature of the transitions is also different: in the case of formula (\ref{08}), it is due to atomic ionization, while the phase transition in Ref.~\cite{Starostin2016} is associated with a change in the dominant neutral component of plasma from H$_2$ to H. However, the shapes of the $P(V)$ curves in Fig.~\ref{fig05} and Fig.~\ref{fig08} are very similar: the same sharp, elongated "beak" for the upper half-wave and smooth, broad lower half-wave. All the features of the $P(V)$ dependencies discussed above are also reproduced in Fig.~\ref{fig08}.

\begin{figure}[]
  \centering
  \includegraphics[width=1.0\linewidth]{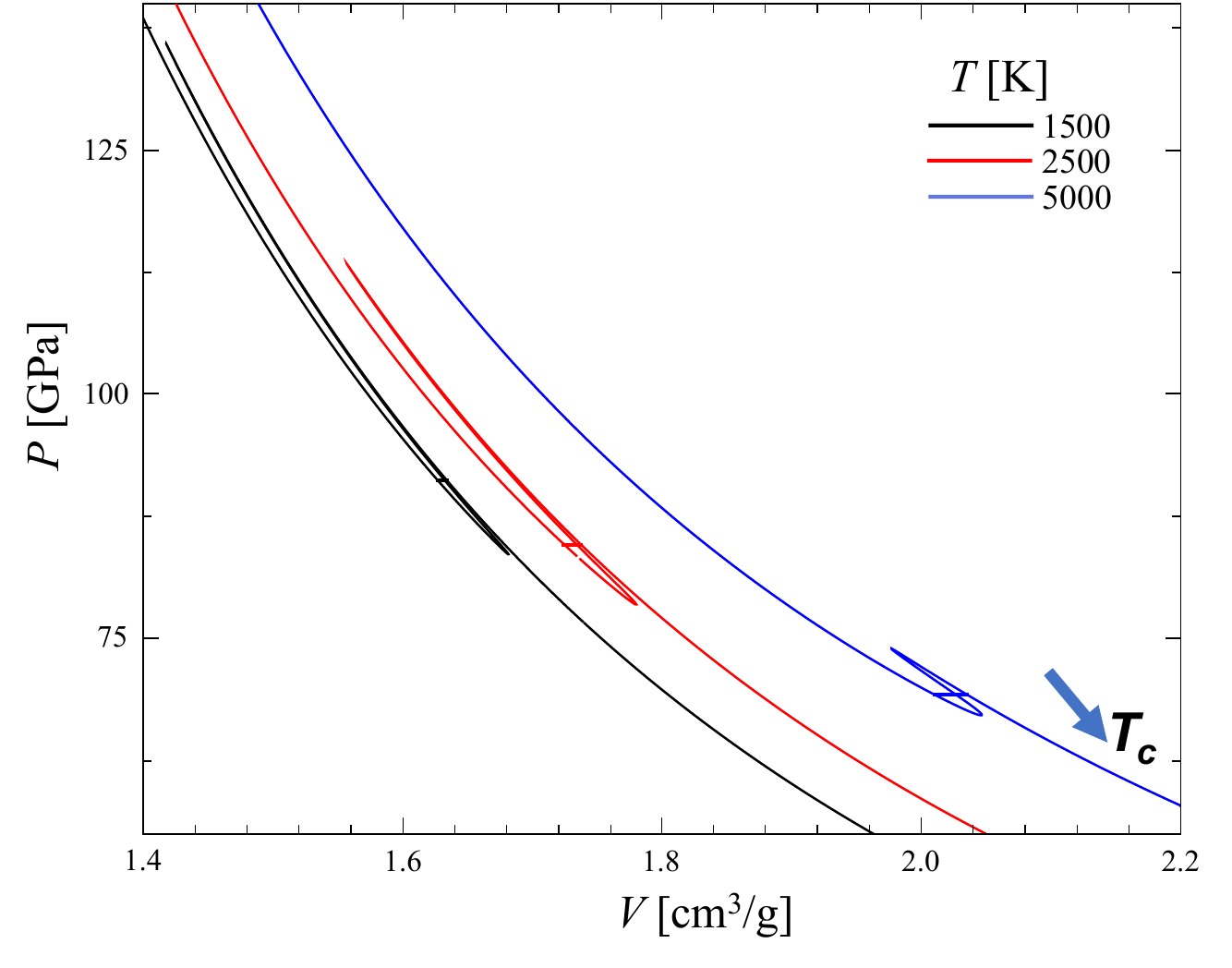}
  \caption{Equation of state $P(V)$ and the phase transition associated with the change of the major ion species from H$_2^+$ to H$^+$ for three temperatures in the current SAHA-D model \cite{Starostin2016}. \footnote{We thank A.V.~Filippov for kindly providing us from his personal archive with this figure, which has not been previously published.}}
  \label{fig09}	
\end{figure}

The authors of Ref.~\cite{Starostin2016} found also another phase transition at large specific volumes, associated with the change of the major ion species from H$_2^+$ to H$^+$. The results for the same temperatures $T=$ 1500, 2500, and 5000~K are shown in Fig.~\ref{fig09} (the $P(V)$ dependence in the phase transition region for $T=5000$~K is highlighted in a separate figure in Ref.~\cite{Starostin2016}). Despite the new and different nature of this phase transition, in this case as well, the upper half-wave turned out to be narrower and sharper than the lower one, and all the features of the $P(V)$ dependencies discussed above are reproduced.

\begin{figure}[]
  \centering
  \includegraphics[width=1.0\linewidth]{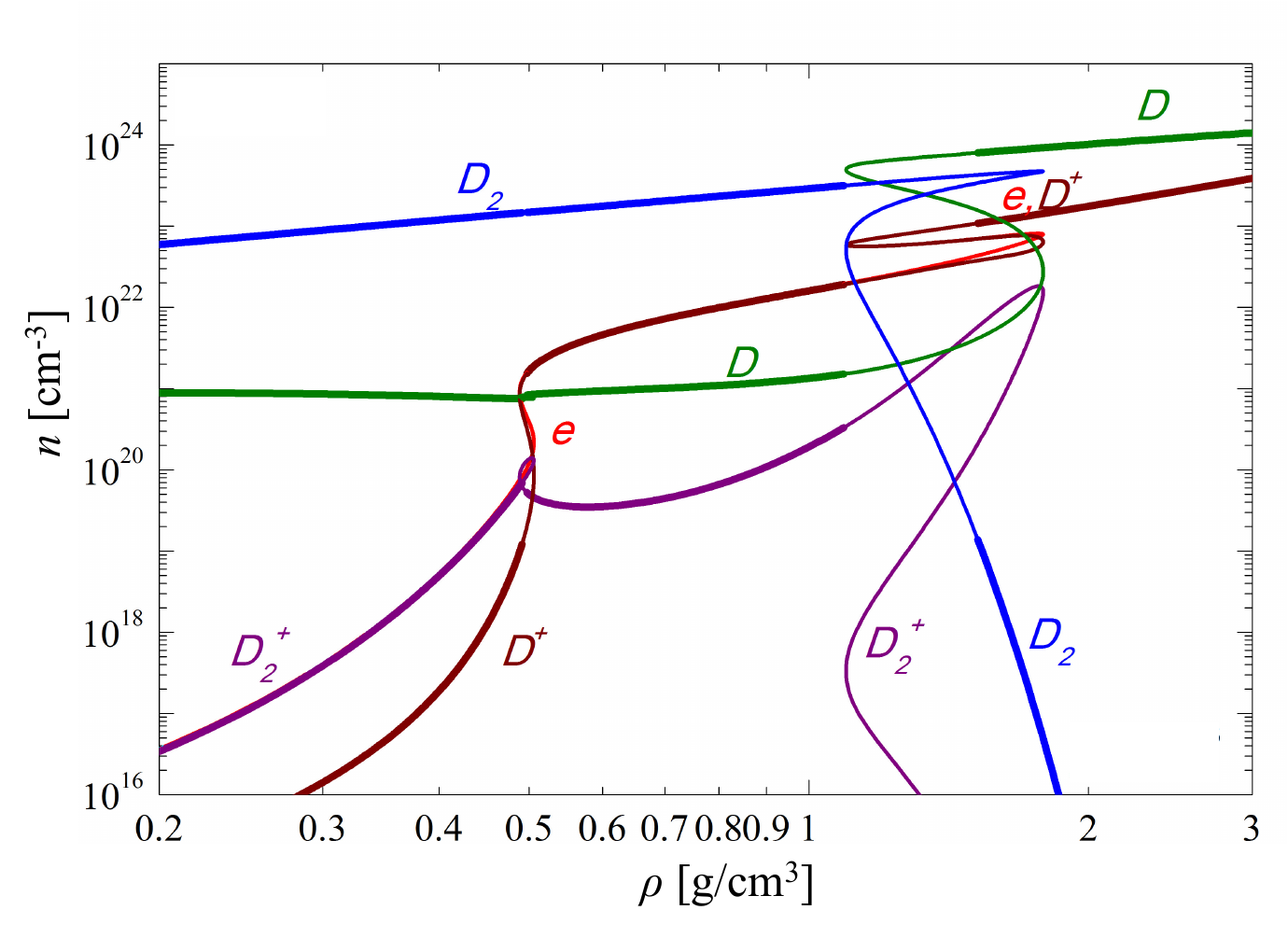}
  \caption{Change in plasma composition with varying density along the $T = 5000$~K isotherm. \footnote{We thank V.K. Gryaznov for kindly providing this previously unpublished figure from his personal archive.}}
  \label{fig10}	
\end{figure}

The plasma composition changes in a rather intricate manner with varying density along the isotherms, due to a whole set of ionization and dissociation reactions. Looking at the 5000~K isotherm shown in Fig.~\ref{fig10}, one can trace where the change of the dominant ion species from H$_2^+$ to H$^+$ occurs, and where the change of the main neutral plasma component from H$_2$ to H takes place.

\begin{figure}[]
  \centering
  \includegraphics[width=1.0\linewidth]{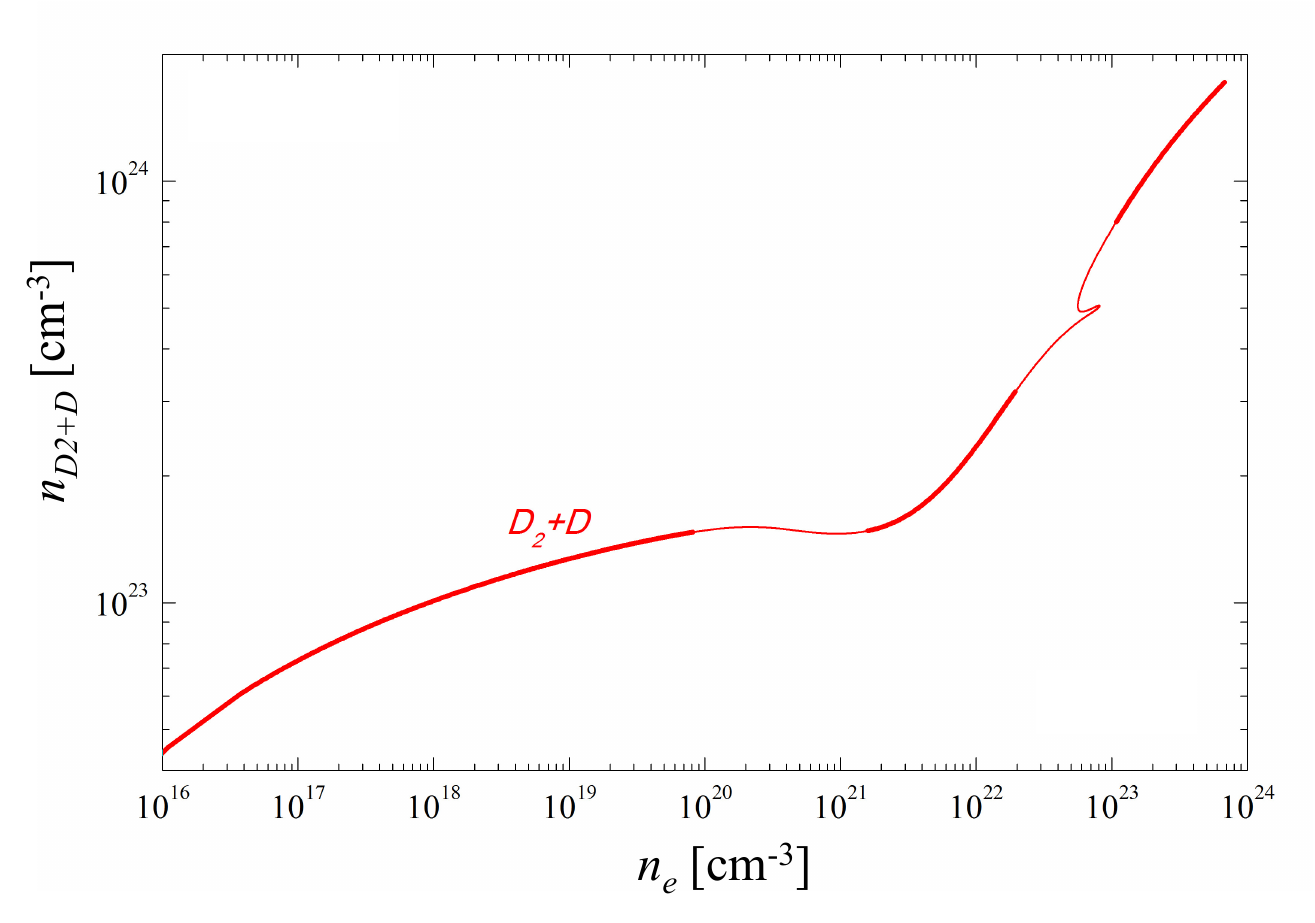}
  \caption{Summary plot: dependence of the total concentration of neutral atoms and neutral molecules on the electron concentration at temperature 5000~K. Two sections of thin red lines correspond to regions where the pressure decreases with increasing density. These regions indicate the presence of two phase transitions. \footnote{We thank V.K. Gryaznov for kindly providing this previously unpublished figure from his personal archive.}}
  \label{fig11}	
\end{figure}

A clearer picture of the phase transitions is shown in Fig.~\ref{fig11}, where for the same temperature of 5000~K the dependence of the total concentration of neutral atoms and neutral molecules on the electron concentration is presented. The figure shows two segments of thin red lines corresponding to two regions where the pressure decreases with increasing density. These are the regions of the two phase transitions. At lower charge concentrations, the transition is associated with a change in the dominant ion species from D$_2^+$ to D$^+$, while at higher concentrations it corresponds to a change in the main neutral component from D$_2$ to D.

Now let us compare Fig.~\ref{fig11} with Fig.~\ref{fig03}. Due to electro-neutrality, the electron concentration is equal to the total concentration of atomic D$^+$ and molecular D$_2^+$ ions. Therefore, both figures show the dependence of the neutral component concentration on the concentration of charged particles. It turns out that in both Fig.~\ref{fig03} and Fig.~\ref{fig11}, the phase transitions correspond to the non-monotonic sections of these dependencies.

At the same time, Fig.~\ref{fig03} is plotted for a three-component plasma, with the neutral component treated as an ideal gas. On the other hand, Fig.~\ref{fig11} is plotted for a five-component plasma, with interactions between all components taken into account and none of them treated as an ideal gas. These differences account for the distinctions between Fig.~\ref{fig11} and Fig.~\ref{fig03}.

At lower charge concentrations in Fig.~\ref{fig11}, the phase transition associated with the change of the main ion species from D$_2^+$ to D$^+$ occurs under conditions that still do not differ significantly from an ideal gas. Therefore, this region in Fig.~\ref{fig11} is close to the corresponding region in Fig.~\ref{fig03}.

At higher charge concentrations in Fig.~\ref{fig11}, the phase transition associated with the change of the main neutral component of the plasma from D$_2$ to D takes place under conditions that are quite far from an ideal gas. Consequently, this region in Fig.~\ref{fig11} differs from Fig.~\ref{fig03}. However, in both Fig.~\ref{fig03} and Fig.~\ref{fig11}, the phase transitions correspond to non-monotonic sections of the dependencies of the neutral component concentration on the charge concentration.

From the series of examples considered, it becomes clear that the determining factor for the existence of the phase transition is the presence of a chemical reaction. In these examples, several different ionization reactions took place. The example where the phase transition is related to the change of the main neutral component from H$_2$ to H indicates that the primary factor for the phase transition is the dissociation reaction, while the Coulomb interaction of charges influences only the properties of the phase transition. In the next section, we consider an example where there is no ionization at all, only dissociation, i.e. the Coulomb interaction is entirely absent.

\section{Phase transition in a model system of diatomic molecules dissociating into atoms}
A completely different chemical model without ionization was applied to warm dense hydrogen by A.L. Khomkin (1945–2022) and A.S. Shumikhin \cite{Khomkin2014,Khomkin2022}. The authors considered a mixture of deuterium atoms and molecules, taking into account the collective cohesion energy of atoms and the pairwise repulsion of neutrals in the Carnahan-Starling hard-sphere approximation. At first glance, this model looks different from those considered in the previous sections of this article. However, there is a fundamental commonality between them, since all models include dissociation/ionization of particles, and particle interactions influence the equilibrium in both approaches.

Indeed, the cohesion energy of atoms provides a weak long-range attraction (analogous to the effective Coulomb attraction in plasma), while the classical Carnahan-Starling repulsion is short-range and strong (similar to the quantum repulsion of electrons from ions in plasma). Therefore, the same analogy with the van der Waals equation works here as in the non-ideal plasma. Therefore, the difference related to different dissociation reactions (electron-ion dissociation in plasma and molecular-atomic dissociation in the model \cite{Khomkin2014,Khomkin2022}) is not essential.

\begin{figure}[]
  \centering
  \includegraphics[width=1.0\linewidth]{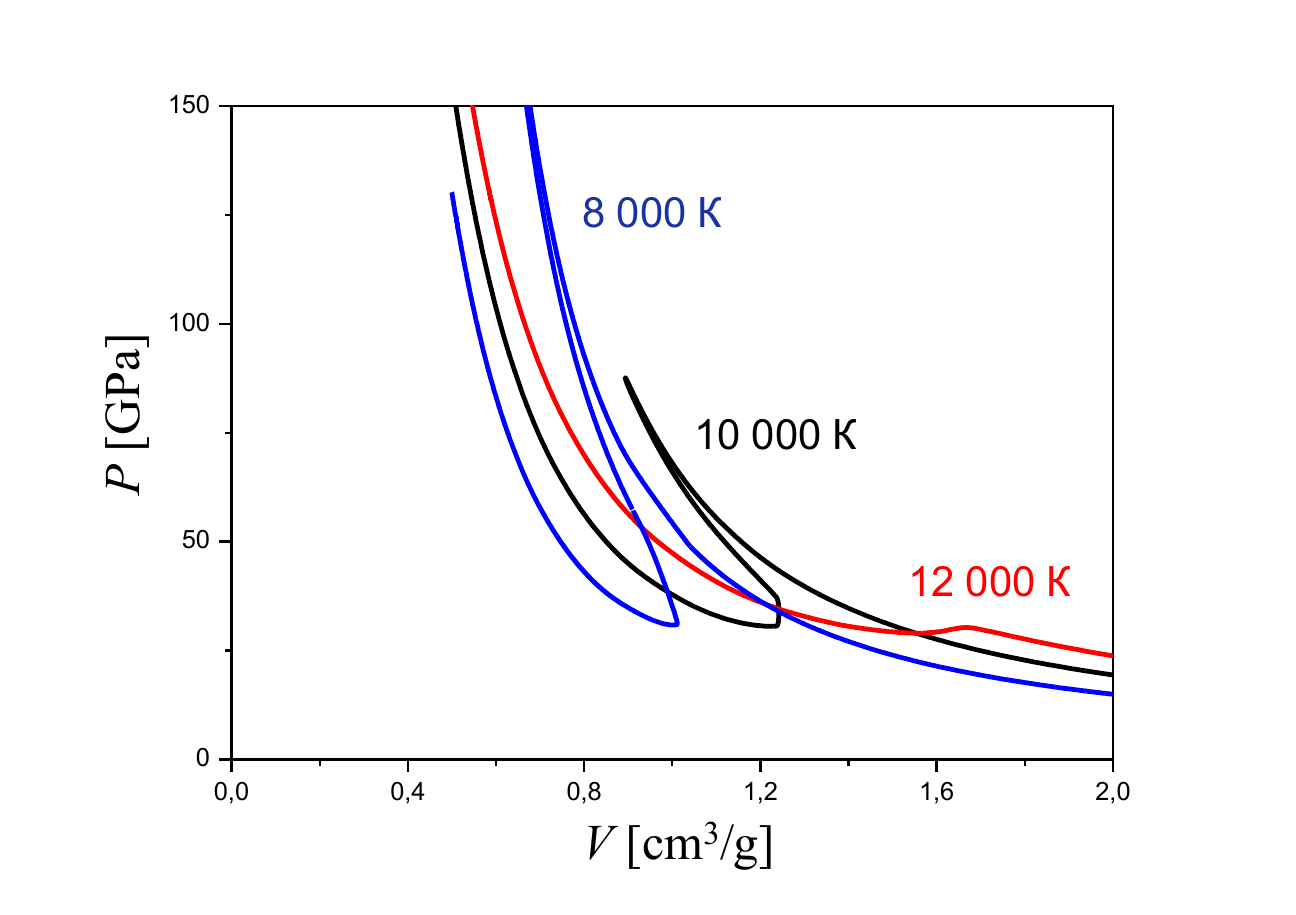}
  \caption{Isotherms $P(V)$ for model warm dense deuterium for three temperature values. \footnote{We thank A.S. Shumikhin for kindly providing us with this previously unpublished figure from his archive.}}
  \label{fig12}	
\end{figure}

\begin{figure}[]
  \centering
  \includegraphics[width=1.0\linewidth]{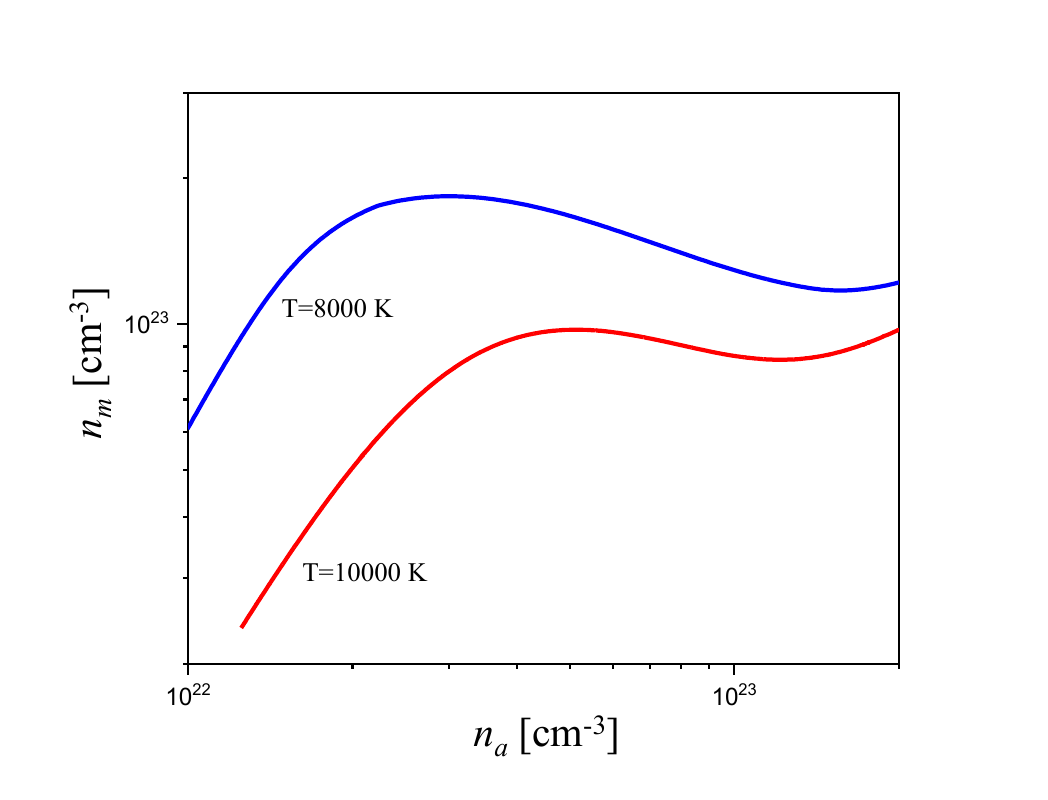}
  \caption{Isotherms $n_m(n_a)$ for model warm dense deuterium at two subcritical temperatures. \footnote{We thank A.S. Shumikhin for kindly providing us with this previously unpublished figure from his archive.}}
  \label{fig13}	
\end{figure}

Consequently, the isotherms $P(V)$ shown in Fig.~\ref{fig12} demonstrate a similar dependence of pressure on specific volume as we analyzed earlier in previous sections. The similarity becomes even more convincing when looking at the isotherms $n_m(n_a)$, showing the dependence of molecular concentration $n_m$ on the number density of atoms $n_a$, presented in Fig.~\ref{fig13} for two subcritical temperatures. These curves have the same character as the $n_a(n_i)$ dependencies for plasma shown in Fig.~\ref{fig03} and in Fig.~\ref{fig06}.

The two phases differ in the degree of molecular dissociation and the concentration of atoms. Each phase is represented in Fig.~\ref{fig12} by its branches of equilibrium and metastable states, limited by spinodal points; however, the points of the phase transition itself were not indicated in \cite{Khomkin2014,Khomkin2022}. Spinodal and binodal points are also not marked in Fig.~\ref{fig13}. Unfortunately, the work was interrupted due to the untimely passing away of A.L.~Khomkin.

The chemical models, whose results are shown in Figs.~\ref{fig02}–\ref{fig09}, \ref{fig12}, and \ref{fig13}, are quite different from each other. Nevertheless, the curve profiles on these figures are quite similar: the same sharp elongated "beak" for the upper half-wave and the smooth broad lower half-wave for the isotherms $P(V)$. This indicates the similarity of the physical nature of these phase transitions, in particular, the presence of chemical ionization or dissociation reactions. Therefore, it is possible to introduce the concept of a new class (type) of first-order phase transitions: ionization or dissociation-driven phase transitions. The proposal of a new class of phase transitions in hydrogen and deuterium involving chemical ionization or dissociation reactions has already been suggested in Ref.~\cite{Khomkin2022}, as the title of that article itself suggests. The authors of Ref.~\cite{Khomkin2022} relied only on their own results and on those taken from the review \cite{Norman1970a}, i.e., essentially on the results of Ref.~\cite{Biberman1969}. As follows from our article, the evidential base turned out to be significantly broader and more convincing.

Note that the model \cite{Khomkin2014,Khomkin2022} has purely methodological interest in connection with the phase transition in fluid hydrogen since it does not describe the jump in electrical conductivity observed in experiments.

\section{Phase transition in fluid hydrogen. Ab initio approach}
\textit{Ab initio} DFT is the approach of a higher level in comparison with chemical models. In this approach, only the total number of electrons and protons/ions is fixed from the start. Atoms, molecules, and other species do not explicitly appear in the calculation but emerge, if needed, only when interpreting the solutions of the Kohn-Sham equations.

Unlike chemical models, which have very weak correlation with experimental data, the \textit{ab initio} approach provides results that can be compared with experiments and are expected to agree with them. To verify the validity of the above-stated hypothesis about a new class (type) of first order ionization/dissociation-driven phase transitions, we proceed to study the properties of fluid hydrogen by quantum molecular dynamics (QMD) methods at temperatures above the melting curve.

\subsection{The method}
Within the framework of the QMD approach, the trajectories of ions $\mathbf{R}(t)$ with masses $M$ are found by solving the classical Newtonian equations of motion
\begin{equation}\label{11}
  M\frac{d^2\mathbf{R}}{dt} = \mathbf{F},
\end{equation}
where the forces $\mathbf{F}$ in the system above are determined based on electronic structure calculations within the framework of the DFT, using the Hellmann–Feynman theorem \cite{feynman1939forces} by the formula
\begin{equation}\label{12}
 \mathbf{F} = -\left( \sum_{n\mathbf{k}}f_{n\mathbf{k}}\langle\phi_{n\mathbf{k}}| \frac{\partial H^{KS}(\mathbf{r},\mathbf{R})}{\partial \mathbf{R}}|\phi_{n\mathbf{k}}\rangle + \frac{\partial U(\mathbf{R}, Z)}{\partial \mathbf{R}}\right).
\end{equation}
The parameters included in the expression (\ref{12}) are as follows: $f_{n\mathbf{k}}$ are the occupancies of the electronic levels, determined by the Fermi-Dirac distribution function; $\phi_{n\mathbf{k}}$ are the single-electron orbitals, which are the solutions of the Kohn-Sham equations \cite{kohn1965self} with the Hamiltonian $H^{KS}(\mathbf{r},\mathbf{R})$, where $\mathbf{r}$ and $\mathbf{R}$ are the coordinates of the electronic and ionic subsystems, respectively; $U(\mathbf{R}, Z)$ is the Coulomb repulsion energy between ions with charge $Z$. The summation is carried out over all electronic states $n$ and $k$ points in the Brillouin zone.

\begin{figure}[]
  \centering
  \includegraphics[width=1.0\linewidth]{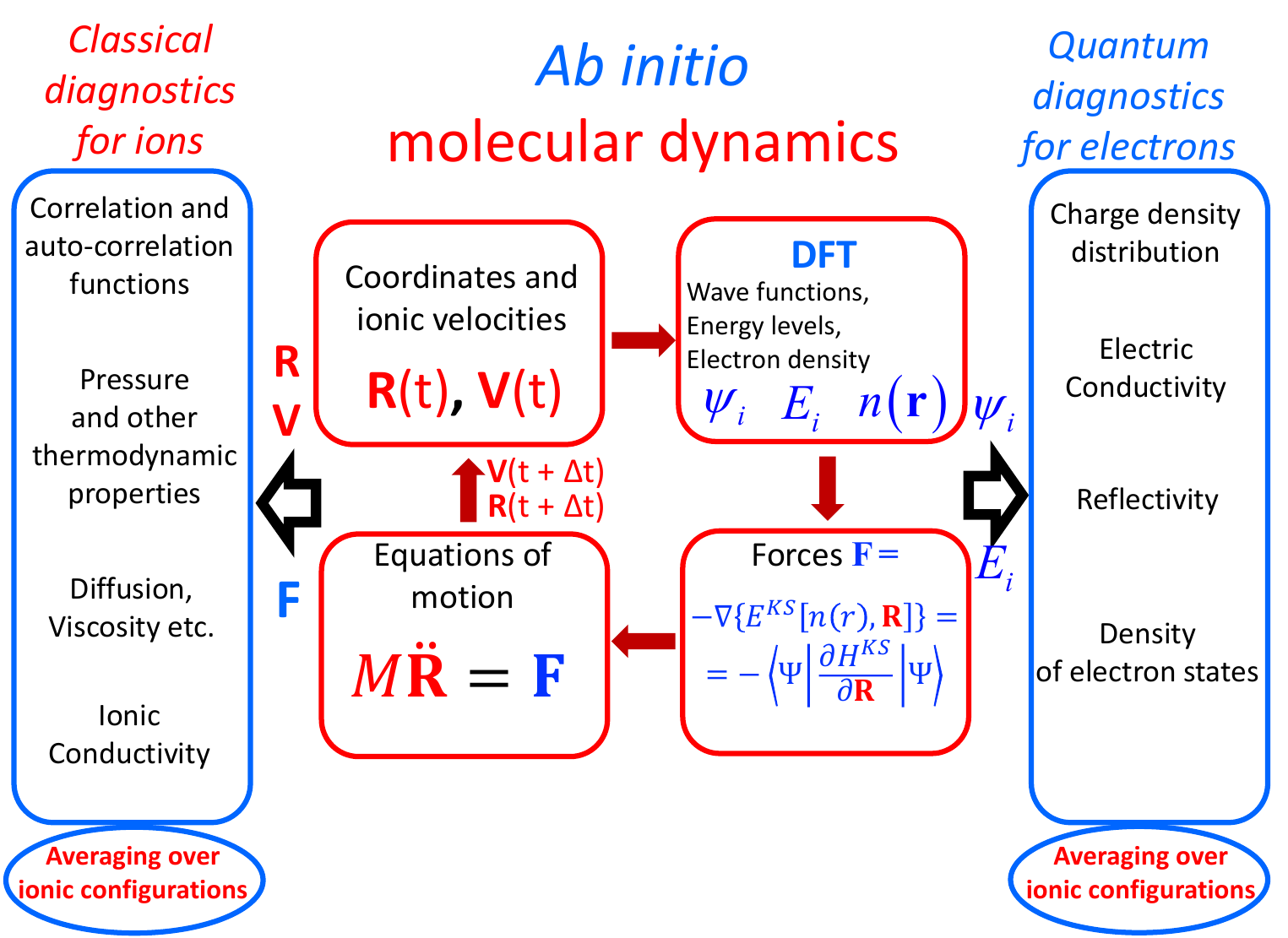}
  \caption{General scheme of the QMD.}
  \label{fig14}	
\end{figure}
The general QMD scheme is shown in Fig.~\ref{fig14}. It contains a central loop in which, at each time step, the coordinates and velocities of ions, the density distribution, wave functions, and electronic energy levels are calculated in parallel. The input parameters are the geometry of the simulation cell, the number of atoms, their initial coordinates $\mathbf{R}(0)$, and velocities $\mathbf{V}(0)$.

For a given atomic configuration under periodic boundary conditions, the solution of the Kohn-Sham equations is determined as a sum of plane waves. For a periodic system, according to Bloch’s theorem, each electron wave function can be represented as
\begin{equation}\label{13}
 \phi_{n\mathbf{k}}(\mathbf{r}) = u_{n\mathbf{k}}(\mathbf{r})e^{i\mathbf{kr}},
\end{equation}
where the function $u_{n\mathbf{k}}(\mathbf{r})$ has the same periodicity as the system under consideration. In the plane wave basis, the function $u_{n\mathbf{k}}(\mathbf{r})$ can be expanded over the reciprocal lattice vectors $\mathbf{G}$
\begin{equation}\label{14}
 u_{n\mathbf{k}}(\mathbf{r}) = \sum_{\mathbf{G}}^{\hbar^2|\mathbf{k} + \mathbf{G}|^2/2m\leq E_{cut}} c_{n\mathbf{k}}(\mathbf{G})e^{i\mathbf{Gr}},
\end{equation}
where $E_{cut}$ is the plane wave basis cutoff energy. The wave vector $\mathbf{k}$ lies within the first Brillouin zone. A finite k-point grid, defined according to the Monkhorst-Pack scheme \cite{Monkhorst1976}, is used to sample the Brillouin zone.

Note that for non-periodic systems, such as liquids, the concept of periodicity is undoubtedly artificial. However, these systems can be treated within the supercell approach, where the corresponding systems are placed into cells containing a large number of particles without internal periodicity, to minimize interactions between repeating images.

In the plane wave basis set, it is quite challenging to describe strongly localized electronic states with rapid oscillations of the wave function at distances close to the atomic core. To solve the problem, we divide electrons into energetically deeper core electrons and valence electrons. We apply the pseudopotential approach within the framework of the projector-augmented wave method \cite{blochl1994projector} in order to take into account the influence of core electrons.

We use the PBE (Perdew–Burke–Ernzerhof) parameterization \cite{Perdew1996} for the exchange-correlation functional. As a solution to the Kohn-Sham equations system, we obtain the electronic density distribution $\rho(\mathbf{r})$, electronic orbitals $\phi_{n\mathbf{k}}(\mathbf{r})$, and their corresponding energy levels $E_{n\mathbf{k}}$. Then, based on the calculation of the system's ground state energy $E_0[\rho(\mathbf{r}), \mathbf{R}]$, we compute the forces between ions according to formula (\ref{12}) and, using these forces, numerically solve the system of equations of motion (\ref{11}).

We perform calculations for the canonical ensemble (NVT). The ion temperature is controlled using the Nosé–Hoover thermostat \cite{Nose1984,Hoover1985}. The electron temperature is equal to the ion temperature and is established as a parameter of the Fermi-Dirac distribution that determines the occupancy of electronic states $f_{n\mathbf{k}}$. This cycle continues until the predetermined number of time steps is completed.

We compute instantaneous values of correlation and autocorrelation functions, pressure, diffusion and viscosity coefficients, and conductivity at each time step, based on the coordinates and velocities of the ions and the forces acting on them. Since the energy functional does not include the kinetic energy of ions, we add a correction in the form of the ideal gas pressure of the ionic component to determine the total pressure
\begin{equation}\label{15}
    P = (\rho_I/\mu)RT + P_{ext},
\end{equation}
where $\rho_I$ is ion mass density, $\mu$ is ion molar mass, $R$ is the gas constant. The external pressure can be determined as the negative value of the average of the diagonal components of the stress tensor
\begin{equation*}
    P_{ext} = -(1/3)\sum_{\alpha}\sigma_{\alpha\alpha}.
\end{equation*}

\subsection{Equation of state}
The equation of state of warm dense hydrogen including the metastable states at the phase transition region within the framework of the QMD approach has been studied in Refs~\cite{Norman2018a,Sartan2019}. The fluid is molecular in the lower density region, while at higher densities the molecules decompose. 

In order to obtain the molecular metastable state, we take the molecular structure for the initial configuration of the QMD run. The first metastable point can be obtained by taking the coordinates and velocities of ions from an equilibrium molecular state and slightly reducing the computational cell size (i.e., increasing the density). The configuration of the new point, brought to equilibrium, can then be used as the starting point for the next metastable state point. In Refs.~\cite{Norman2018a,Sartan2019}, the molecular metastable branches of isotherms have been obtained by gradually increasing the density by 0.5–2.0\% and relaxing each new configuration.

To maintain a metastable molecular state, we perform the calculations in the microcanonical ensemble (NVE), i.e., with the thermostat turned off. The thermostat should be used only to relax the initial configuration and then switched off for the calculation of metastable states of the molecular fluid. The thermostat acts as an additional perturbation that reduces the lifetime of metastable states.

The temperature drift is insignificant for the NVE trajectories. The loss of metastability usually occurs after a change in the cell volume, since this also acts as a perturbation. In such cases, we chose a different initial configuration. The perturbation may also cause temperature drift. This issue can be resolved by turning the thermostat on for a short time (less than 0.5~ps) or by selecting another initial configuration. Hydrogen may still remain metastable with the NVT simulation run, but the probability of losing metastability and the transition to the ionized atomic-molecular stable state increases.

Approaching the spinodal endpoints along isotherms is a common procedure in simulations of Lennard-Jones systems and other systems where interparticle forces are computed as derivatives of the potential energy with respect to distance. However, in the current approach, forces are calculated via the Hellmann-Feynman theorem at each step of numerically integrating Newton’s equations. This procedure introduces additional perturbations that can limit the modeling of short-lived metastable states.

The lifetime of metastable states exceeds 15~ps near the two-phase region and is less than 5~ps at the boundary of the metastable region at high densities.

\begin{figure}
    \centering
    \includegraphics[width=0.8\linewidth]{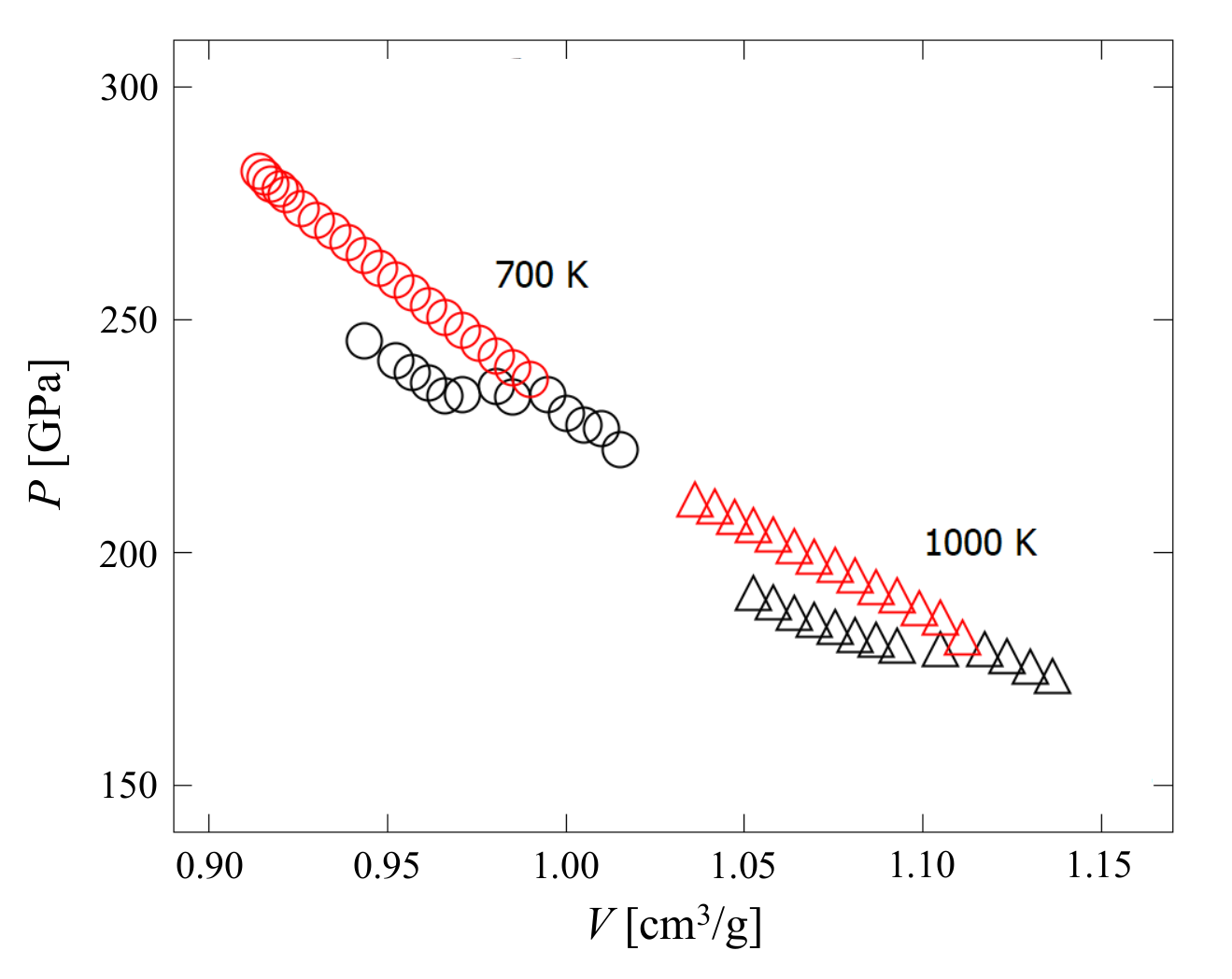}
    \caption{Isotherms of warm dense hydrogen at 700 and 1000~K \cite{Norman2018a, Sartan2019}. The metastable branch of the molecular phase is highlighted in red for each temperature.}
    \label{fig15}
\end{figure}

\begin{figure}
    \centering
    \includegraphics[width=0.8\linewidth]{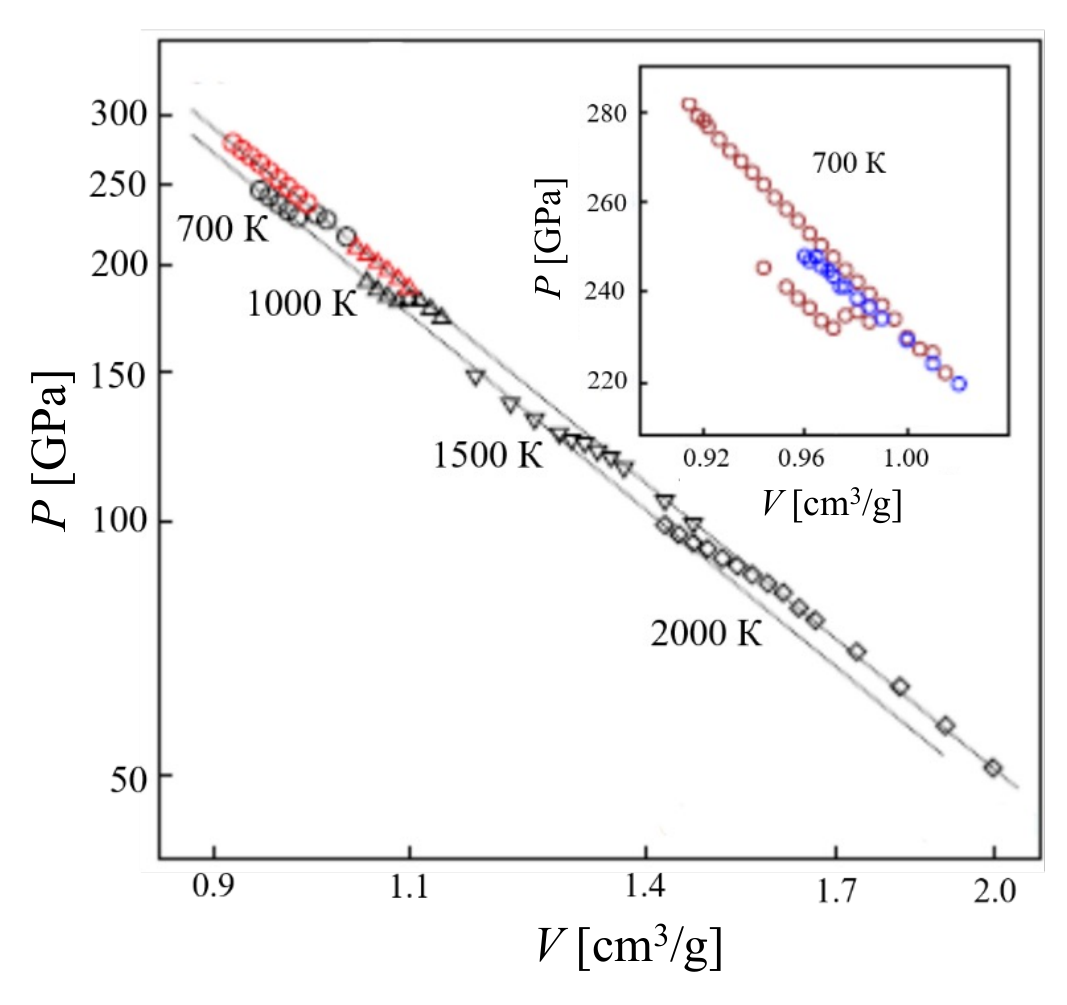}
    \caption{Equation of state of warm dense hydrogen from Refs.~\cite{Norman2018a,Sartan2019} in a double logarithmic scale. Metastable states are highlighted in red. The inset shows the results of hydrogen isotherm calculations at 700~K for different numbers of particles: 512 (red circles) and 1024 (blue circles).}
    \label{fig16}
\end{figure}

The results for the 700 and 1000~K isotherms \cite{Norman2018a,Sartan2019}, including the metastable branch on which hydrogen remains molecular as the density increases, are shown in Fig.~\ref{fig15}. The pressure difference between metastable and stable branches at a given temperature is 15~GPa. Four isotherms (1500 and 2000~K in addition to the previous two) in a double-logarithmic scale are shown in Fig.~\ref{fig16}. The method and calculation parameters for the stable branches of the isotherms presented in Figs.~\ref{fig15} and \ref{fig16} are identical to those used in work \cite{Lorenzen2010}. However, in the article \cite{Lorenzen2010} metastable states have been hardly studied; only the initial points of the metastable branches on the 700~K isotherm near the jump in specific volume were obtained.

The size of the metastable region depends on the number of particles (inset in Fig.~\ref{fig16}). The lifetime of the metastable state decreases as the number of atoms increases, since a larger volume increases the probability of nucleation centers of the high-density phase appearing.

According to the data shown in Fig.~\ref{fig15}, one can notice that the modulus of the derivative of pressure on volume doesn't decrease with the increasing density along the metastable branches. The absence of isotherm flattening indicates the existence of metastable states at higher pressures, closer to the spinodal. These states could not be obtained for the reasons mentioned above.

Metastable states and density jumps are absent on the isotherms at 1500 and 2000~K. However, as it is shown in Refs~\cite{Lorenzen2010,Norman2018a,Sartan2019}, the increase in conductivity by four orders of magnitude is observed at 1500~K within a relatively narrow density range. The isotherms of equilibrium and metastable states of the same phase lie along the same lines for different temperatures (dashed lines in Fig.~\ref{fig16}). These lines coincide with the binodal of the subcritical temperature, and the pressure difference between them decreases exponentially in absolute value with increasing temperature. The proximity and almost the parallelism (in a double logarithmic scale) of the two dashed lines maintain the uncertainty of the critical temperature within several thousand kelvin.

The metastable isotherms obtained using the QMD approach are in qualitative agreement with the predictions reported in Ref.~\cite{Biberman1969} (Fig.~\ref{fig04}). Notable features include a relatively small density discontinuity and substantial overlap between the metastable and equilibrium branches. A comparable result is produced by a modern chemical plasma model that accounts for hydrogen molecule ionization \cite{Starostin2016}, discussed earlier in Section III (see Figs.~\ref{fig08} and \ref{fig09}). The original isotherms from the chemical model with a simple Padé approximation, shown in Fig.~\ref{fig05}, also display a profile similar to that of the QMD results. The fluid hydrogen isotherms obtained from QMD and chemical plasma models qualitatively match the schematic depiction in Fig.~\ref{fig01}(b). The phase coexistence line for fluid hydrogen appears as a long, curved, and very narrow tongue, making it challenging to clearly define its boundaries and to precisely determine the location of the critical point $T_c$. 

The phase coexistence line terminates at the critical point on the phase diagram, and various parameters change continuously beyond the two-phase region. However, parameters may also exhibit anomalies in the supercritical region. For example, extrema can occur in the second derivatives of the Gibbs thermodynamic potential, such as the compressibility, the thermal expansion coefficient, the heat capacity, or other functions \cite{brazhkin2011widom,brazhkin2012fomin,brazhkin2014true}. Such anomalies in the pressure–temperature plane start from the critical point, extend the phase equilibrium line, and form the so-called Widom line. The Widom line depends on the parameter chosen, and a series of such lines can be constructed. For hydrogen, in Refs.~\cite{Sartan2019,Norman2021}, the Widom line has been determined using the minima of the pressure derivatives with respect to temperature along isochores.

Estimates of the critical temperature for warm dense hydrogen vary significantly across theoretical studies, even when the same computational method is applied: 1500~K in Ref.~\cite{Lorenzen2010} and 4000~K in Ref.~\cite{Norman2017,Norman2017a}. An estimate of 1000~K was predicted for a chemical model in the earlier work \cite{Norman1968}. The results shown in Fig.~\ref{fig16} are consistent with this range. However, the accuracy of all available approaches is insufficient to determine the exact value of the critical temperature since the indeterminate intermediate region between the Widom line and the phase equilibrium curve is rather broad.

\section{Plasma nature of the phase transition in warm dense hydrogen}
\begin{figure}
    \centering
    \includegraphics[width=1\linewidth]{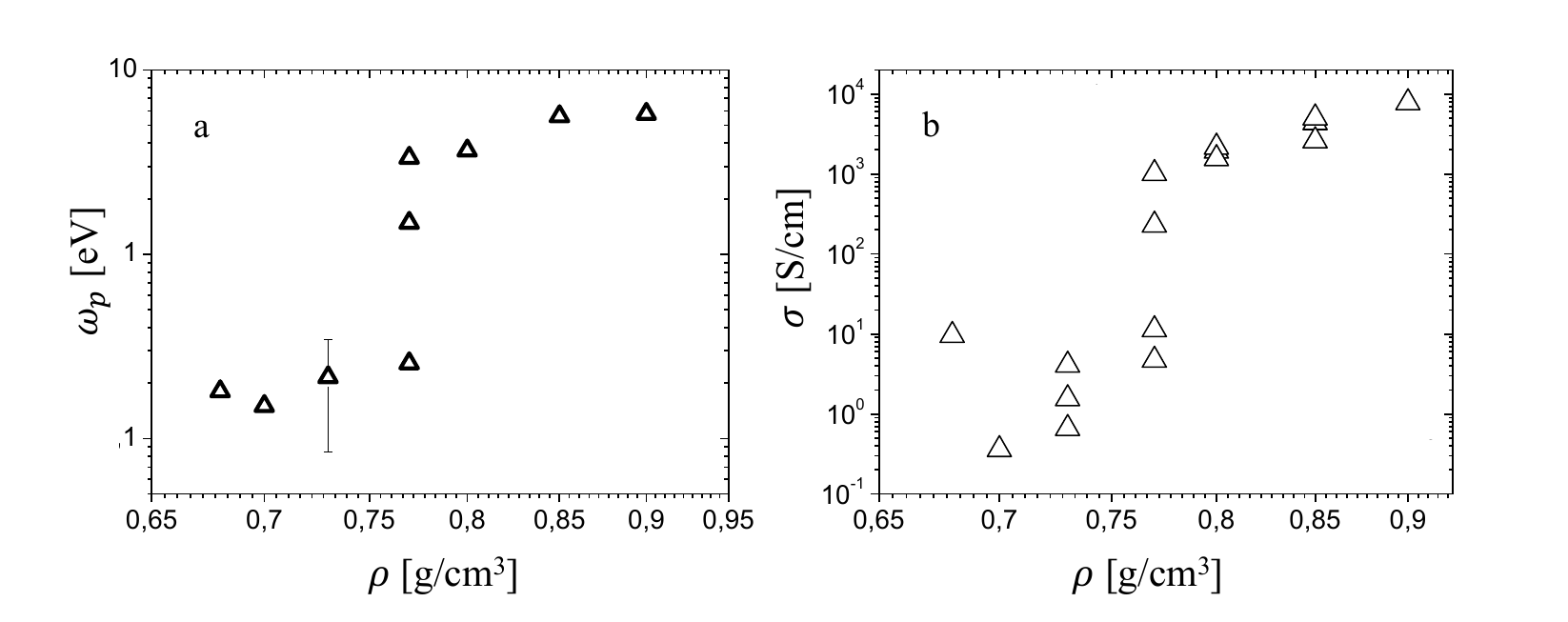}
    \caption{Dependence of the plasma frequency (a) and electrical conductivity (b) on the density of hot dense hydrogen at $T=1500$~K. In panel (a), the statistical error in determining the plasma frequency is shown for a density of 0.73~g/cm$^3$. The $\omega_p$ values are averaged over a set of configurations.}
    \label{fig:pf}
\end{figure}

According to the definition of plasma proposed in Refs.~\cite{1979MAtomA,kamenetskii1972plasma}, one of the principal characteristics of a plasma is the existence of plasma oscillations. Therefore, the discontinuity in the plasma frequency provides additional evidence for the plasma nature of the phase transition. A sharp increase in the plasma frequency $\omega_p$ is observed in Ref.~\cite{norman2015plasma} within the same narrow density range as the jump in electrical conductivity $\sigma$. An example of the dependence of the plasma frequency and electrical conductivity on density at 1500~K, obtained earlier in work~\cite{norman2015plasma}, is presented in Fig.~\ref{fig:pf}. The procedures used to determine these quantities are described in Ref.~\cite{norman2015plasma}.

The similarity between the QMD results \cite{Norman2018a, Sartan2019} and the results of the chemical model \cite{Biberman1969, Starostin2016} indicates the plasma nature of the fluid–fluid phase transition in warm dense hydrogen. Here we can identify three features, which indicate direct evidence of the PPT.

\begin{enumerate}
    \item Significant overlap between the metastable and stable branches of the isotherm, resulting in a three-valued region in the pressure–density dependence. This feature, predicted within the chemical model for a singly ionized plasma in Ref.~\cite{Biberman1969}, also appears in the QMD simulation results for warm dense hydrogen in Refs.~\cite{Norman2017, Norman2017a, Norman2018a, Sartan2019}. Here we should emphasize that the existence of metastable states is direct evidence of a first-order phase transition. This feature is important in particular for determining the nature of the formation of the conducting phase of fluid hydrogen, since the theoretically predicted density jump is quite small.

    \item A sharp increase in electrical conductivity in a narrow density range along with a density jump on the isotherm. This is accompanied by partial dissociation and ionization of hydrogen molecules. Comparison of simulation results with experimental data indicates that the location of the fluid-fluid phase transition in hydrogen can be associated with the semiconductor–dielectric (or semimetal–dielectric) boundary \cite{Norman2021}. The subsequent formation of a metallic atomic fluid is the crossover.

    \item Along with the increase in electrical conductivity, a sharp rise in the plasma frequency is also observed in warm dense hydrogen. This indicates an increase in the free electron number density. Thus, under the studied density and temperature conditions, warm dense hydrogen can be considered as a plasma. The form of the plasma frequency versus density dependence provides additional evidence of the ionization mechanism of the phase transition.
\end{enumerate}

\section{Conclusions}
Let us highlight three conclusions.

1. The phase diagram presented in Ref.~\cite{Biberman1969} is fundamentally different from the van der Waals phase diagram. In the present work, we have analyzed the results obtained in Refs.~\cite{Gryaznov2009, Starostin2016, Khomkin2014, Khomkin2022}, as well as results obtained by the authors of the present study in particular for this analysis. Within the framework of three different chemical models including a multicomponent plasma model \cite{Gryaznov2009, Starostin2016}, a dissociating dense gas of diatomic molecules without ionization \cite{Khomkin2014, Khomkin2022}, and the three-component plasma model used in this work, we confirm the universal nature of the prediction \cite{Biberman1969}. Therefore, we found three characteristic features of the isotherms $P(V)$, i.e., the dependence of pressure $P$ on specific volume $V$ for warm dense matter or a nonideal plasma:
\begin{enumerate}[(a)]
    \item the overlap of the metastable branch of one phase with the metastable and equilibrium branches of another phase within a certain range of specific volumes,
    \item due to the triple-valued nature of $P(V)$ in a certain range of specific volumes, there exists an isolated segment of metastable states in this range,
    \item the pressure of the phase transition decreases as the critical temperature is approached, i.e., as the temperature increases. In other words, the phase equilibrium curve in $P$–$T$ coordinates has a negative slope.
\end{enumerate}

\begin{figure}
    \centering
    \includegraphics[width=1\linewidth]{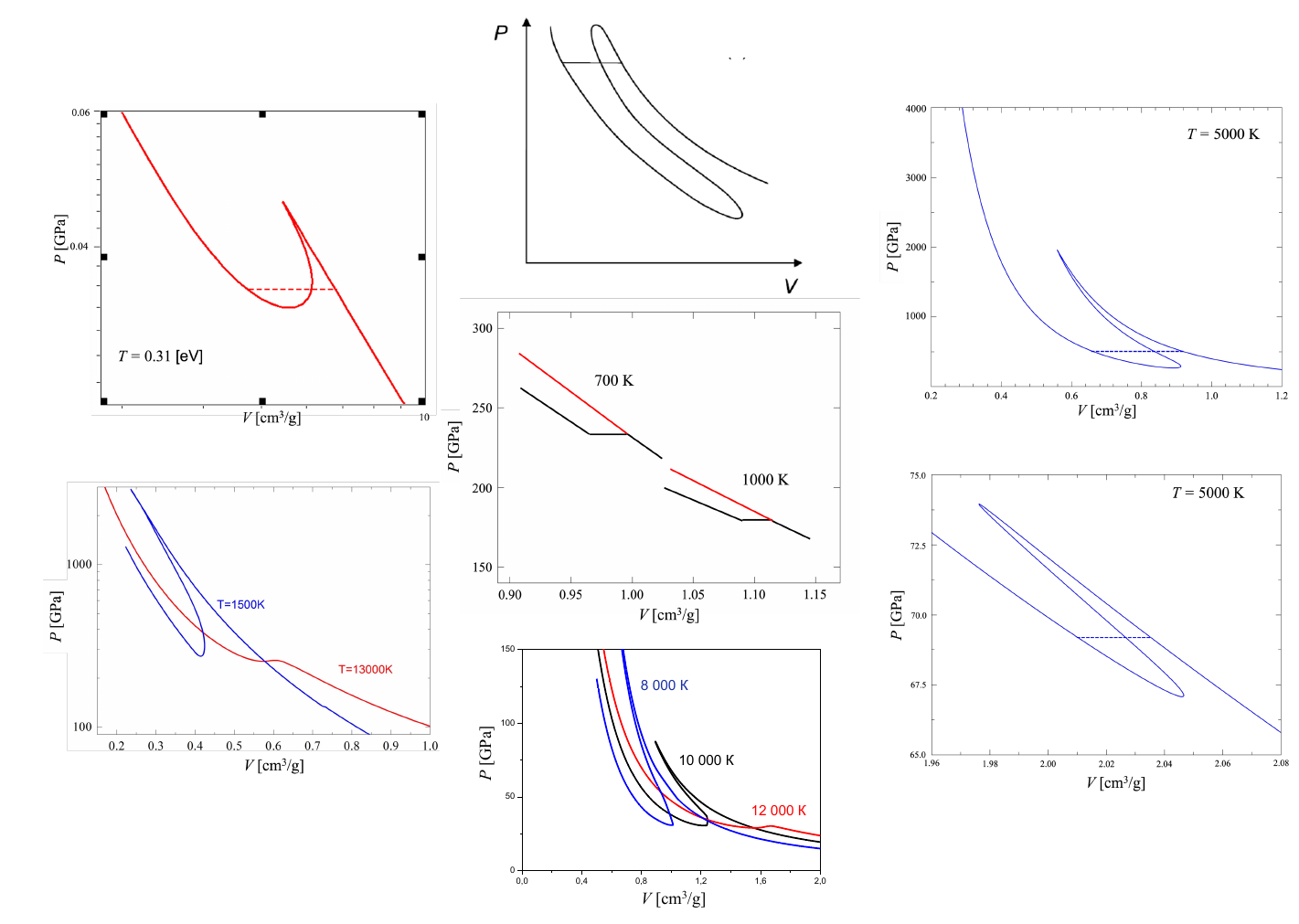}
    \caption{A summary figure that brings together all the results discussed in this article: six $P(V)$ isotherms obtained from various chemical models are arranged around the circle, and in the center is placed an example of a $P(V)$ isotherm obtained within the QMD.}
    \label{fig_final}
\end{figure}

These conclusions, presented graphically in Fig.~\ref{fig_final}, emphasize their universality: we deliberately do not specify in the caption of Fig.~\ref{fig_final} which of the $P(V)$ isotherms corresponds to which chemical model, since without such clarification they are difficult to distinguish.

Regarding the feature (c), it is observed in all the theoretical approaches considered that describe the equation of state of warm dense hydrogen in the region of formation of the conducting state. However, in the general case, according to the analysis of the model equation of state for the PPT given in the original works \cite{Norman1968, Biberman1969, Norman1970a, Norman1970}, the slope may also be positive. In particular, Ref.~\cite{dharma2025ionic} shows that the phase transition in warm dense silicon, which has a mechanism similar to the PPT, exhibits a positive slope of the phase equilibrium curve.

2. We identify the same features using the results from Refs.~\cite{Norman2017,Norman2017a,Norman2018a,Sartan2019,Norman2021} along with additional calculations within the framework of QMD. Features (a) and (c) are confirmed, while the second feature (b) has so far been observed only within chemical models. The quantum calculation results do not rule out this possibility, although direct confirmation has not yet been obtained.

These features sharply distinguish the obtained $P(V)$ isotherms (Fig.~\ref{fig_final}) from van der Waals isotherms and other models. A common physical nature of the transitions, which is present in all examples, both classical and one quantum, is the presence of a jump in ionization degree (i.e., a discontinuous change in ionization) or a jump in dissociation degree (i.e., a discontinuous change in molecular dissociation). Notably, a jump in electrical conductivity is not a mandatory signature, since the dissociative phase transition \cite{Khomkin2014,Khomkin2022} occurs in a model substance where electrical conductivity is absent altogether. All this provides grounds to single out a new class (type) of first-order phase transitions: ionization or dissociation-driven phase transitions. Plasma phase transition is a particular case of this class.

3. The fluid–fluid phase transition in warm dense hydrogen, observed experimentally as well as in chemical models and quantum calculations, should be classified as a plasma phase transition. First, its isotherms have the same form as those characteristic of ionization or dissociation-driven phase transitions. Second, the transition is accompanied by a jump in electrical conductivity, observed both experimentally and theoretically; moreover, both phases are conductive. Third, plasma oscillations exist in both phases, and the transition is accompanied by a jump in the plasma frequency.

\begin{acknowledgements}
    The authors express their deep gratitude to V.K. Gryaznov, A.V. Filippov, and A.S. Shumikhina for kindly providing us from their archives with figures that have not been published anywhere before. G.N. is supported by the Ministry of Science and Higher Education of the Russian Federation (State Assignment No. 075-01129-23-00). I.S. is supported by the European Union - NextGenerationEU under the Italian Ministry of University and Research (MUR) projects PRIN2022 PNRR-P2022MC742PNRR, CUP E53D23018440001.
\end{acknowledgements}

\bibliographystyle{apsrev4-2.bst}
\bibliography{refs}

\end{document}